\documentclass[pra, twocolumn, amsmath, amssymb, superscriptaddress]{revtex4-1}

\usepackage{graphicx, braket, xcolor, mathrsfs, float}

\newcommand{\+}{^\dagger}
\newcommand{\ha}{\hat a}
\newcommand{\hb}{\hat b}
\newcommand{\hc}{\hat c}
\newcommand{\vx}{\vec{x}}
\newcommand{\vy}{\vec{y}}
\newcommand{\hvX}{\hat{\vec{X}}}
\newcommand{\hvY}{\hat{\vec{Y}}}
\newcommand{\lrvec}[1]{\overleftrightarrow{#1}}

\newcommand{\Rev}[1]{{\color{black} #1}}

\begin{document}

\title{Driven-dissipative four-mode squeezing of multilevel atoms in an optical cavity}

\author{Bhuvanesh Sundar$^*$}
\affiliation{Center for Theory of Quantum Matter, University of Colorado, Boulder, CO 80309, USA}
\affiliation{JILA, NIST, Department of Physics, University of Colorado, Boulder, CO 80309, USA}
\thanks{Now at Rigetti Computing, Berkeley, CA 94710, USA}
\author{Diego Barberena}
\affiliation{Center for Theory of Quantum Matter, University of Colorado, Boulder, CO 80309, USA}
\affiliation{JILA, NIST, Department of Physics, University of Colorado, Boulder, CO 80309, USA}
\author{Ana Maria Rey}
\affiliation{Center for Theory of Quantum Matter, University of Colorado, Boulder, CO 80309, USA}
\affiliation{JILA, NIST, Department of Physics, University of Colorado, Boulder, CO 80309, USA}
\author{Asier Pi\~neiro Orioli}
\affiliation{Center for Theory of Quantum Matter, University of Colorado, Boulder, CO 80309, USA}
\affiliation{JILA, NIST, Department of Physics, University of Colorado, Boulder, CO 80309, USA}

\begin{abstract}
We utilize multilevel atoms trapped in a driven resonant optical cavity to produce scalable multi-mode squeezed states for quantum sensing and metrology. While superradiance or collective dissipative emission by itself has been typically a detrimental effect for entanglement generation in optical cavities, in the presence of additional drives it can also be used as an entanglement resource. In a recent work~\cite{sundar2023squeezing}, we described a protocol for the dissipative generation of two-mode squeezing in the dark state of a six-level system with only one relevant polarization. There we showed that up to two quadratures can be squeezed. Here, we develop a generalized analytic treatment to calculate the squeezing in any multilevel system where atoms can collectively decay by emitting light into two polarization modes in a cavity. We show that in this more general system up to four spin squeezed quadratures can be obtained. We study how finite-size effects constrain the reachable squeezing, and analytically compute the scaling with $N$. 
Our findings are readily testable in current optical cavity experiments with alkaline-earth-like atoms.

\end{abstract}

\maketitle
\section{Introduction}
Creating many-body states of matter with large useful entanglement that can be harnessed for quantum sensing and metrology is a highly sought-after goal. 
Optical cavities are natural candidates for creating such types of entangled states since photon-mediated interactions between atoms allow for the generation of collective (i.e., fully symmetric) quantum many-body states with entanglement that grows with the atom number $N$. One particular type of entangled state that can be created in this way is the spin squeezed state~\cite{ma2011quantum, schleier2010squeezing, leroux2010implementation, chen2014cavity,Cox2016,hosten2016measurement,Pezze2018}, i.e.,~a state with a reduced variance along some spin direction. 

Most of the effort so far has been focused on the generation of squeezing by restricting the dynamics to two levels per atom~\cite{Pezze2018}, using either coherent interactions~\cite{kitagawa1993squeezed, borregaard2017one, braverman2019near, hu2017vacuum, RLS_TSS_2018}, or dissipation~\cite{dalla2013dissipative, muschik2011dissipatively, kastoryano2011dissipative, diehl2008quantum, verstraete2009quantum}. However, the use of the full multilevel atomic structure can open up new opportunities for creating different types of collective entangled states~\cite{vitagliano2011spin, vitagliano2014spin}, such as multimode squeezed states, i.e.,~states with two or more squeezed spin directions. Multimode squeezed states are not easily accessible in collective two-level systems, and they could be useful for multi-parameter estimation~\cite{demkowicz2020multi}.

In Ref.~\cite{sundar2023squeezing}, we proposed to use coherent driving and superradiance on multilevel systems with one relevant cavity polarization as a resource for generation of scalable two-mode squeezing. We also showed ways to store squeezed states in dark manifolds that are robust to collective dissipation ~\cite{sundar2023squeezing,orioli2022emergent}.
In this paper, we describe the dissipative squeezing dynamics for a wide range of multilevel structures in the case of two relevant cavity polarizations. We derive the condition for the system to be stable to quantum fluctuations, and show that up to four spin variables are typically squeezed in this more general system. We study how finite-size effects constrain the reachable squeezing, and analytically compute the scaling of the squeezing with $N$.

The paper is outlined as follows. In Sec.~\ref{sec: implementation}, we describe the proposed experimental setup and derive the effective master equation. In Sec.~\ref{sec: mf}, we describe the mean-field (MF) physics and stability to quantum fluctuations. In Sec.~\ref{sec: correlations}, we calculate the quantum correlations that develop between the atoms, and the emergent squeezing, during the driven-dissipative dynamics. In Sec.~\ref{sec: two-level} and~\ref{sec: multilevel}, we apply the techniques developed in prior sections to an effective two-level and multilevel system, respectively. We conclude in Sec.~\ref{sec: discussion}.

\begin{figure*}
\includegraphics[width=2.0\columnwidth]{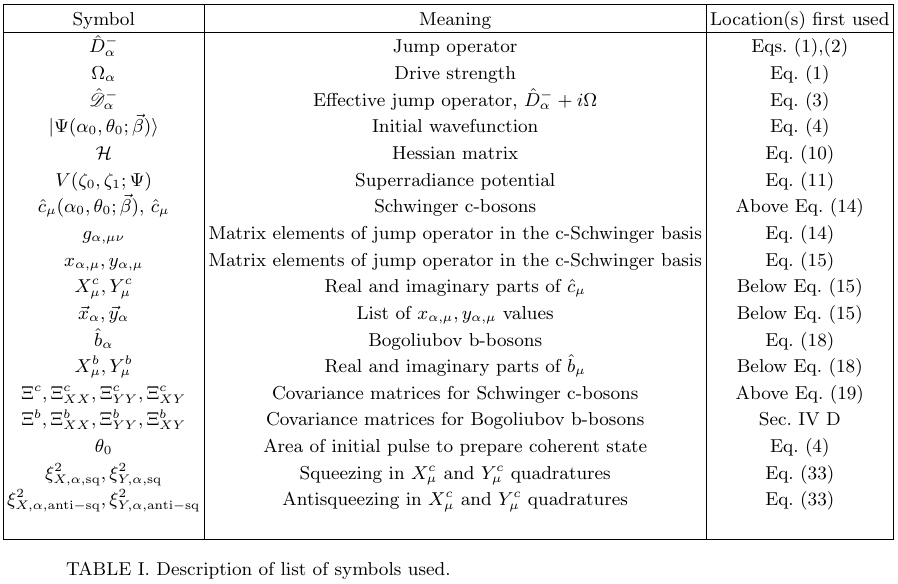}
\label{table}
\end{figure*}

We note that we include a reference table with all symbols used in Table~\ref{table} that the reader might find helpful.

\section{System and initial state}\label{sec: implementation}
\begin{figure}[t]
\includegraphics[width=1.0\columnwidth]{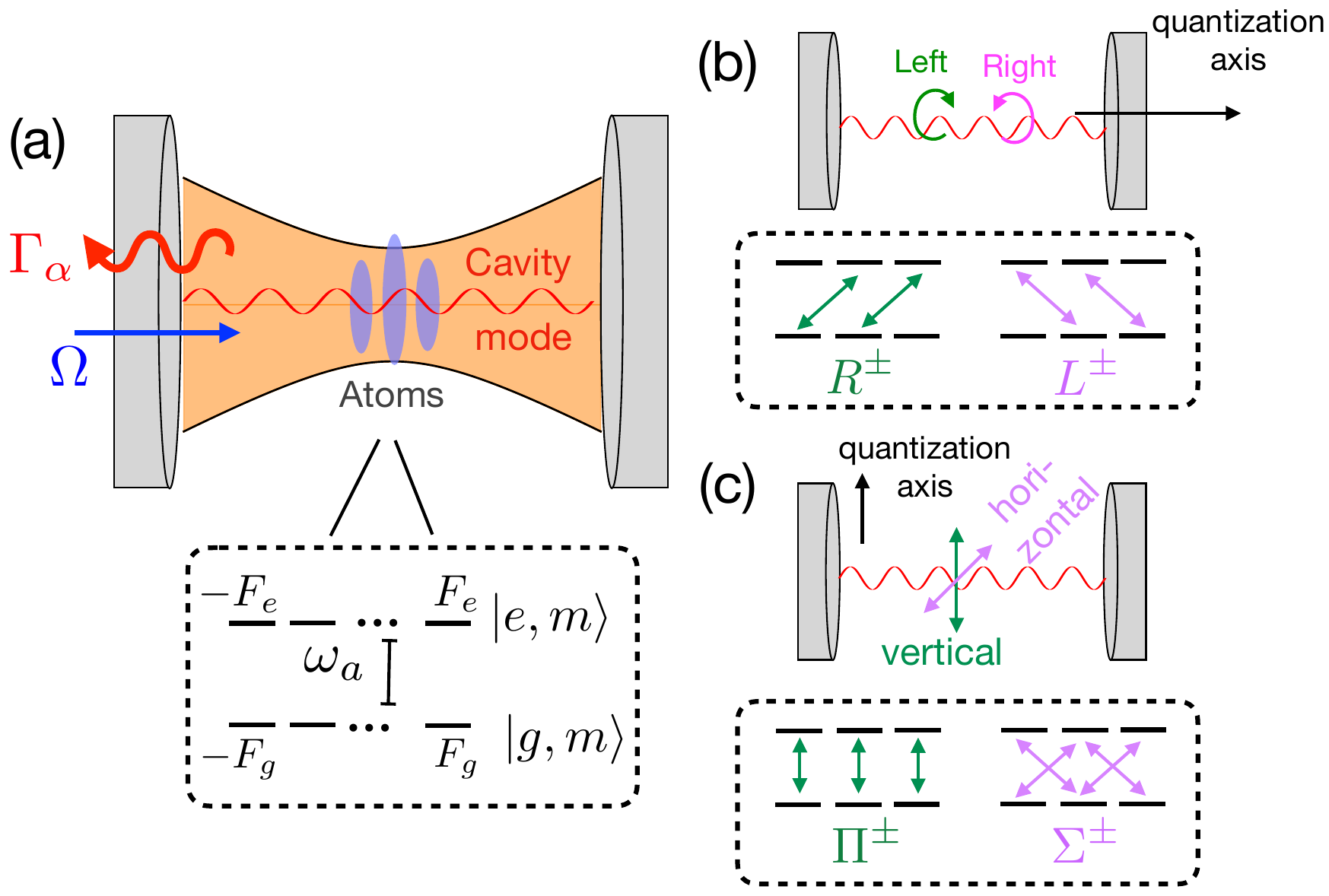}
\caption{(a) An ensemble of atoms trapped in a deep lattice in an optical cavity, with the cavity frequency on resonance with the atomic transition from the ground states to the excited states (with spins $F_g$ and $F_e$), $\omega_c = \omega_a$. The cavity is driven by a resonant laser, and the atoms superradiantly decay at rate $\Gamma$ from the excited to the ground states. (b,c) Transitions driven by the collective atomic excitation operators. (b) illustrates the $L^\pm$ and $R^\pm$ transitions due to coupling to a left or right circularly polarized photon when we choose the quantization axis to be parallel to the cavity axis. (c) illustrates the $\Pi^\pm$ and $\Sigma^\pm$ transitions due to coupling to a vertically or horizontally polarized photon when we choose the quantization axis to be perpendicular to the cavity axis.}
\label{fig: setup}
\end{figure}

We consider an ensemble of $N$ multilevel atoms pinned in a deep optical lattice within an optical cavity [see Fig.~\ref{fig: setup}(a)]. We consider the atoms to have a degenerate ground manifold with $2F_g+1$ levels, labeled $\ket{g,m} (-F_g \leq m \leq F_g)$, and a long-lived degenerate excited manifold with $2F_e+1$ levels labeled $\ket{e,m} (-F_e \leq m \leq F_e)$. Here, $F_g$ and $F_e$ are the spins in the ground and excited manifolds, and $m$ denotes the angular momentum projection along the quantization axis. The ground-excited transition frequency is $\omega \equiv \omega_a$. 

The cavity is assumed to be resonant with the atomic transition and to support a pair of photon modes with degenerate angular frequency $\omega_c = \omega_a = \omega$ and orthogonal polarizations [see Figs.~\ref{fig: setup}(b-c)], both perpendicular to the cavity axis [see Figs.~\ref{fig: setup}(b-c)]. The atoms couple to these two cavity modes with single-photon Rabi frequency $2g$. The cavity modes are also driven with a resonant laser with frequency, $\omega_l = \omega$. Additionally, photons can leak out of the cavity at a rate $\kappa$.

If $\kappa \gg g \sqrt{N}$, we can adiabatically eliminate the cavity photons and obtain an effective master equation for the atoms only, $\hbar\frac{d\rho}{dt} = -i[\hat H_{\rm drive}, \rho] + \mathcal{L}[\rho]$. The effective Hamiltonian and dissipation terms for the atoms are
\begin{align} \label{eqn: H}
&\hat H_{\rm drive} = \sum_\alpha \frac{\hbar\Omega_\alpha}{2} (\hat{D}_\alpha^- + \hat{D}_\alpha^+), \\
\label{eqn: L}
&\mathcal{L}[\rho] = \sum_\alpha \hbar\Gamma \left(\hat{D}_\alpha^- \rho \hat{D}_\alpha^+ - \frac{1}{2}\hat{D}_\alpha^+\hat{D}_\alpha^- \rho - \frac{1}{2}\rho\hat{D}_\alpha^+\hat{D}_\alpha^-\right),
\end{align}
where $\Omega_\alpha$ is the intracavity drive strength, and $\hat{D}_\alpha^+$ is a collective atomic operator that excites atoms by absorbing an $\alpha$-polarized photon. If the $\alpha$-polarized photon has angular momentum projection $l_\alpha=\pm 1,0$ along the quantization axis, then $\hat D_\alpha^+ = \sum_i \hat d_{i,\alpha}^+$ with $i$ running over the atoms, and $\hat d^+_{i,\alpha} = \sum_{m} C^{m}_\alpha \hat s^+_{m,i,\alpha}$, where the sum runs over the ground state atomic levels. The single-particle spin-raising operator $\hat{s}^+_{m,i,\alpha} = \ket{e,m+ l_\alpha}_i \bra{g,m}_i$ drives a transition between the levels $\ket{g,m}$ and $\ket{e,m+l_\alpha}$, and $C^{m}_\alpha = \langle F_g,m; 1,l_\alpha | F_e, m+l_\alpha \rangle$ is the Clebsch-Gordan coefficient for this transition. 
$\Gamma =4 g^2/\kappa$ is the cavity-induced decay of an atom from the excited manifold.

The master equation $\hbar\frac{d\rho}{dt} = -i[\hat H_{\rm drive}, \rho] + \mathcal{L}[\rho]$ can also be written as
\begin{equation}\label{eqn: Leff}
\hbar\frac{d\rho}{dt} = \mathcal{L}'[\rho] \equiv \sum_\alpha \hbar\Gamma \left(\hat{\mathscr{D}}_\alpha^- \rho \hat{\mathscr{D}}_\alpha^+ - \frac{1}{2}\hat{\mathscr{D}}_\alpha^+\hat{\mathscr{D}}_\alpha^- \rho - \frac{1}{2}\rho\hat{\mathscr{D}}_\alpha^+\hat{\mathscr{D}}_\alpha^-\right)
\end{equation}
where $\hat{\mathscr{D}}_\alpha^- = \hat{D}_\alpha^- + i\Omega_\alpha/\Gamma$. 
Detailed derivations of Eqs.~\eqref{eqn: H},~\eqref{eqn: L}, and~\eqref{eqn: Leff} are given in Appendix~\ref{app: effective master equation}.

The basis states $\ket{g,m}$ and $\ket{e,m}$ are associated with a particular choice of the quantization axis. In this paper, we will either choose the quantization axis to be along the cavity axis, or perpendicular to the cavity axis, and we will explicitly specify this where necessary. Similarly, the cavity supports two polarizations of light, which we will choose to decompose into either linear modes or circular modes. Whenever we choose the atomic quantization axis along the cavity axis, we will choose the polarizations as left-handed (denoted $\alpha = L$ and having $l_\alpha = -1$) and right-handed (denoted $\alpha = R$ and having $l_\alpha = +1$) [Fig.~\ref{fig: setup}(b)]. 
Whenever we choose the atomic quantization axis to be perpendicular to the cavity axis, we will define the polarizations as vertical (denoted $\alpha = \Pi$ and having $l_\alpha = 0$) and horizontal (denoted $\alpha = \Sigma$, which includes both $l_\alpha = 1$ and $-1$) [Fig.~\ref{fig: setup}(c)] and define $\hat D^+_\Sigma = (\hat D^+_L + \hat D^+_R)/\sqrt{2}$.

We initialize the atoms in a product of single-particle ground states $\ket{G_{\vec{\beta}}} = \sum_m \beta_m \ket{g,m}$, and apply a laser pulse of duration $\tau$ and polarization $\alpha_0$ such that $\hat H_{\rm drive}\tau/\hbar = \theta_0 \hat D^x_{\alpha_0}$. This leaves the atoms in the coherent state
\begin{equation}\label{eqn: Psi}
\ket{\Psi(\alpha_0,\theta_0; \vec{\beta} )} = \exp(-i \theta_0 \hat D^x_{\alpha_0}) \ket{G_{\vec{\beta}}}^{\otimes N},
\end{equation}
where $\hat D^x_{\alpha} = (\hat D^+_\alpha + \hat D^-_\alpha)/2$, and $\hat D^y_\alpha = (\hat D^+_\alpha - \hat D^-_\alpha)/2i$. We will denote the polarization that is orthogonal to $\alpha_0$ as $\alpha_1$.

The goal of this paper is to study the properties of the system at the steady state, i.e.,~$\mathcal{L}'[\rho_{\rm ss}] = 0$.
In general, our multilevel system contains a continuum of steady states, since the steady state realized by the dynamics as $t\rightarrow\infty$ depends on the choice of initial state and the parameter $\Omega_\alpha/N\Gamma$.
To constrain the number of possibilities, we will only consider initial states $\ket{\Psi(\alpha_0,\theta_0; \vec{\beta} )}$ as given in Eq.~(\ref{eqn: Psi}) for which the single-particle observables $\braket{\hat{O}}$ are approximately stationary from the beginning and focus on the behavior of the fluctuations captured by higher-order observables.

\section{Mean-field approximation}\label{sec: mf}
We discuss first the properties of the steady state in a MF approximation. For collective systems, MF assumes $\braket{\hat{O}_1 \hat{O}_2} \approx \braket{\hat{O}_1}\braket{\hat{O}_2}$ for any set of collective single-body spin operators $\hat{O}_1$ and $\hat{O}_2$ \footnote{By collective single-body spin operator, we mean an operator of the form $\hat{O} = \sum_i \hat{o}_i$, where $\hat{o}_i$ is an operator that acts only on atom $i$.}.
This approximation works well when $N$ is large and can be seen as the leading-order expansion in powers of $1/N$.
Under this approximation, the master equation [Eq.~\eqref{eqn: Leff}] for any collective single-body spin variable $\braket{\hat O}$ reduces to
\begin{align}\label{eqn: O mf}
\frac{d}{dt}\braket{\hat O}_{\rm MF} \approx & \frac{\Gamma}{2}\sum_\alpha \braket{\hat{\mathscr{D}}^+_\alpha}_{\rm MF} \braket{ [\hat O, \hat{\mathscr{D}}^-_\alpha] }_{\rm MF} 
\nonumber\\&
+ \braket{ [\hat{\mathscr{D}}^+_\alpha, \hat O] }_{\rm MF}\braket{ \hat{\mathscr{D}}^-_\alpha}_{\rm MF} .
\end{align}
Here, we used that the commutator $[\hat{\mathscr{D}}^\pm_\alpha, \hat O]$ is a collective single-body spin operator.

\subsection{Mean-field stationary state}

A sufficient condition for making all spin variables stationary at the mean-field level is to choose the drive $\Omega_\alpha$ and the initial state $\ket{\Psi(\alpha_0,\theta_0;\vec{\beta})}$ such that $\braket{\Psi(\alpha_0,\theta_0;\vec{\beta}) \vert \hat{\mathscr{D}}^-_\alpha \vert \Psi(\alpha_0,\theta_0;\vec{\beta})} = 0$ [see Eq.~\eqref{eqn: O mf}]. Satisfying this requires two conditions:
\begin{align}\label{eqn: mf steady state}
& \braket{\Psi(\alpha_0,\theta_0;\vec{\beta}) \vert \hat{D}^x_\alpha \vert \Psi(\alpha_0,\theta_0;\vec{\beta})} = 0, \\
\label{eqn: Omega steady state}
& \braket{\Psi(\alpha_0,\theta_0;\vec{\beta}) \vert \hat{D}^y_\alpha \vert \Psi(\alpha_0,\theta_0;\vec{\beta})} = \Omega_\alpha/\Gamma.
\end{align}
Throughout this paper, we will choose $\ket{\Psi(\alpha_0,\theta_0;\vec{\beta})}$ and $\Omega_\alpha$ to satisfy Eqs.~\eqref{eqn: mf steady state} and~\eqref{eqn: Omega steady state}, i.e.,~we only consider initial states that are stationary states at the mean-field level.
Moreover, we will later consider only examples where the continuous drive has the same polarization as the preparation pulse, i.e.,~$\Omega_{\alpha_1}=0$, but the discussion in the following sections does not assume this.

\subsection{Stability of the mean-field state}
The mean-field stationary state may be stable or unstable to quantum fluctuations. If it is stable, the fluctuations remain small and the mean-field state $\rho_{\rm MF} = \ket{\Psi(\alpha_0,\theta_0;\vec{\beta})} \bra{\Psi(\alpha_0,\theta_0;\vec{\beta})}$ turns out to be a good approximation to the full quantum steady state $\rho_{\rm ss}$, which satisfies~\cite{puri1979exact, somech2022quantum, barberena2019driven}
\begin{equation}\label{eqn: steady state cond}
\hat{\mathscr{D}}^+_\alpha \hat{\mathscr{D}}^-_\alpha \rho_{\rm ss} \approx 0.
\end{equation}
Note that the approximate sign `$\approx$' means that the above expression is zero up to higher-order corrections in $1/N$ which are qualitatively irrelevant for our purposes.
In our analytical approximation, we have $\braket{ \hat{\mathscr{D}}^+_\alpha \hat{\mathscr{D}}^-_\alpha } = 0$ at the steady state, as is shown below.

In the stable phase, the dynamics of $\braket{ \hat{\mathscr{D}}^+_\alpha \hat{\mathscr{D}}^-_\alpha }$ is well captured by making an approximation where we set the third-order cumulant to zero,
$\braket{\hat{O}_1 \hat{O}_2}\braket{\hat{O}_3} + \braket{\hat{O}_1 \hat{O}_3}\braket{\hat{O}_2} + \braket{\hat{O}_2 \hat{O}_3}\braket{\hat{O}_1} - 2\braket{\hat{O}_1}\braket{\hat{O}_2}\braket{\hat{O}_3} - \braket{\hat{O}_1 \hat{O}_2 \hat{O}_3} \approx 0$.
Under this approximation and further assuming the mean-field stationary condition [Eqs.~\eqref{eqn: mf steady state} and~\eqref{eqn: Omega steady state}] is also met, the master equations for $\braket{ \hat{\mathscr{D}}^+_{\alpha_0} \hat{\mathscr{D}}^-_{\alpha_0} }$ and $\braket{ \hat{\mathscr{D}}^+_{\alpha_1} \hat{\mathscr{D}}^-_{\alpha_1} }$ couple to the equations for $\braket{ \hat{\mathscr{D}}^+_{\alpha_0} \hat{\mathscr{D}}^-_{\alpha_1} }$ and $\braket{ \hat{\mathscr{D}}^+_{\alpha_1} \hat{\mathscr{D}}^-_{\alpha_0} }$. The coupled equations are
\begin{widetext}
\begin{equation}\label{eqn: eqns in cumulant approx}
\partial_t \left(\begin{array}{c} \braket{\hat{\mathscr{D}}^+_{\alpha_0}\hat{\mathscr{D}}^-_{\alpha_0}} \\ \braket{\hat{\mathscr{D}}^+_{\alpha_0}\hat{\mathscr{D}}^-_{\alpha_1}} \\ \braket{\hat{\mathscr{D}}^+_{\alpha_1}\hat{\mathscr{D}}^-_{\alpha_0}} \\ \braket{\hat{\mathscr{D}}^+_{\alpha_1}\hat{\mathscr{D}}^-_{\alpha_1}} \end{array}\right)
\approx 
-\Gamma\left(\begin{array}{cccc} 
\lambda_{00} & \frac{\lambda_{01}}{2} & \frac{\lambda_{10}}{2} & 0 \\
\frac{\lambda_{10}}{2} & \frac{\lambda_{00}+\lambda_{11}}{2} & 0 & \frac{\lambda_{10}}{2}\\
\frac{\lambda_{01}}{2} & 0 & \frac{\lambda_{00}+\lambda_{11}}{2} & \frac{\lambda_{01}}{2}\\
0 & \frac{\lambda_{01}}{2} & \frac{\lambda_{10}}{2} & \lambda_{11}
\end{array}\right)
\left(\begin{array}{c} \braket{\hat{\mathscr{D}}^+_{\alpha_0}\hat{\mathscr{D}}^-_{\alpha_0}} \\ \braket{\hat{\mathscr{D}}^+_{\alpha_0}\hat{\mathscr{D}}^-_{\alpha_1}} \\ \braket{\hat{\mathscr{D}}^+_{\alpha_1}\hat{\mathscr{D}}^-_{\alpha_0}} \\ \braket{\hat{\mathscr{D}}^+_{\alpha_1}\hat{\mathscr{D}}^-_{\alpha_1}} \end{array}\right),
\end{equation}
\end{widetext}
where we denoted $\lambda_{ij} = \braket{[\hat{\mathscr{D}}^-_{\alpha_i}, \hat{\mathscr{D}}^+_{\alpha_j}]}$. The value of $\lambda_{ij}$ is a constant at leading order in $N$, $\lambda_{ij} \approx \braket{[\hat{\mathscr{D}}^-_{\alpha_i}, \hat{\mathscr{D}}^+_{\alpha_j}]}_{\rm MF}$, since it is a single-particle observable and is therefore stationary. Note that Eq.~\eqref{eqn: eqns in cumulant approx} is independent of $\Omega$, which means that the stability is determined by the light emission properties alone.

Generically, demanding that $\braket{ \hat{\mathscr{D}}^+_{\alpha_0} \hat{\mathscr{D}}^-_{\alpha_0} }$ and $\braket{ \hat{\mathscr{D}}^+_{\alpha_1} \hat{\mathscr{D}}^-_{\alpha_1} }$ decay to zero, Eq.~\eqref{eqn: eqns in cumulant approx} requires that $\braket{ \hat{\mathscr{D}}^+_{\alpha_0} \hat{\mathscr{D}}^-_{\alpha_1} }$ and $\braket{ \hat{\mathscr{D}}^+_{\alpha_1} \hat{\mathscr{D}}^-_{\alpha_0} }$ also decay to zero, since they are coupled. This means that the matrix in Eq.~\eqref{eqn: eqns in cumulant approx} needs to be positive definite. \Rev{The condition for positive definiteness of this matrix can be obtained by rewriting Eq.~\eqref{eqn: eqns in cumulant approx} in terms of the sums and differences of correlators:
\begin{widetext}
\begin{equation}
\partial_t
\left(\begin{array}{c}
\braket{\hat{\mathscr{D}}^+_{\alpha_0} \hat{\mathscr{D}}^-_{\alpha_0}} + \braket{\hat{\mathscr{D}}^+_{\alpha_1} \hat{\mathscr{D}}^-_{\alpha_1}}\\
\braket{\hat{\mathscr{D}}^+_{\alpha_0} \hat{\mathscr{D}}^-_{\alpha_1}} + \braket{\hat{\mathscr{D}}^+_{\alpha_1} \hat{\mathscr{D}}^-_{\alpha_0}}\\
\braket{\hat{\mathscr{D}}^+_{\alpha_0} \hat{\mathscr{D}}^-_{\alpha_0}} - \braket{\hat{\mathscr{D}}^+_{\alpha_1} \hat{\mathscr{D}}^-_{\alpha_1}}\\
\braket{\hat{\mathscr{D}}^+_{\alpha_0} \hat{\mathscr{D}}^-_{\alpha_1}} - \braket{\hat{\mathscr{D}}^+_{\alpha_1} \hat{\mathscr{D}}^-_{\alpha_0}}
\end{array}\right)
= -\frac{\Gamma M}{2}
\left(\begin{array}{c}
\braket{\hat{\mathscr{D}}^+_{\alpha_0} \hat{\mathscr{D}}^-_{\alpha_0}} + \braket{\hat{\mathscr{D}}^+_{\alpha_1} \hat{\mathscr{D}}^-_{\alpha_1}}\\
\braket{\hat{\mathscr{D}}^+_{\alpha_0} \hat{\mathscr{D}}^-_{\alpha_1}} + \braket{\hat{\mathscr{D}}^+_{\alpha_1} \hat{\mathscr{D}}^-_{\alpha_0}}\\
\braket{\hat{\mathscr{D}}^+_{\alpha_0} \hat{\mathscr{D}}^-_{\alpha_0}} - \braket{\hat{\mathscr{D}}^+_{\alpha_1} \hat{\mathscr{D}}^-_{\alpha_1}}\\
\braket{\hat{\mathscr{D}}^+_{\alpha_0} \hat{\mathscr{D}}^-_{\alpha_1}} - \braket{\hat{\mathscr{D}}^+_{\alpha_1} \hat{\mathscr{D}}^-_{\alpha_0}}
\end{array}\right)
\end{equation}
where
\begin{equation}
M = \left(\begin{array}{cccc}
\lambda_{00}+\lambda_{11} &
\lambda_{01}+\lambda_{10} &
\lambda_{00}-\lambda_{11} & \lambda_{01}-\lambda_{10} \\
\lambda_{01}+\lambda_{10} & 
\lambda_{00}+\lambda_{11} & 0 & 0\\
\lambda_{00}-\lambda_{11} & 0 &
\lambda_{00}+\lambda_{11} & 
0\\
\lambda_{10}-\lambda_{01} & 0 & 0 &
\lambda_{00}+\lambda_{11}
\end{array}\right),
\end{equation}
\end{widetext}
and $\lambda_{ij} = \braket{[\hat{\mathscr{D}}^-_{\alpha_i}, \hat{\mathscr{D}}^+_{\alpha_j}]}$. 
The correlations $\braket{\hat{\mathscr{D}}^+_{\alpha} \hat{\mathscr{D}}^-_{\beta}}$ decay to zero if the real parts of the eigenvalues of $M$ are all positive. This occurs if the symmetric part of $M$ is positive definite. The eigenvalues of the symmetric part, $(M+M^T)/2$, are $\lambda_{00}+\lambda_{11}$, $\lambda_{00}+\lambda_{11}$, and $\lambda_{00}+\lambda_{11} \pm \sqrt{ \left(\lambda_{00} - \lambda_{11} \right)^2 + \left(\lambda_{01}+\lambda_{10} \right)^2 }$. All four eigenvalues will be positive if the latter two are positive, and the condition for the latter two eigenvalues being positive is equivalent to requiring}
\begin{equation}\label{eqn: stability cond}
\mathcal{H} = \left( \begin{array}{cc}
\lambda_{00} & \frac{ \lambda_{01}+\lambda_{10} }{2} \\
\frac{ \lambda_{01}+\lambda_{10} }{2} & \lambda_{11}
\end{array}\right) \succ 0.
\end{equation}

The time scale of the dynamics due to Eq.~\eqref{eqn: eqns in cumulant approx} is $O(1/N\Gamma)$. Corrections beyond the cumulant approximation drive dynamics on a time scale of $O(1/\sqrt{N}\Gamma) \gg 1/N\Gamma$ [see Appendix~\ref{app subsec: higher-order effect on stability}].

\subsection{Superradiance potential}\label{subsec: potential}
\begin{figure}[t]
\includegraphics[width=\columnwidth]{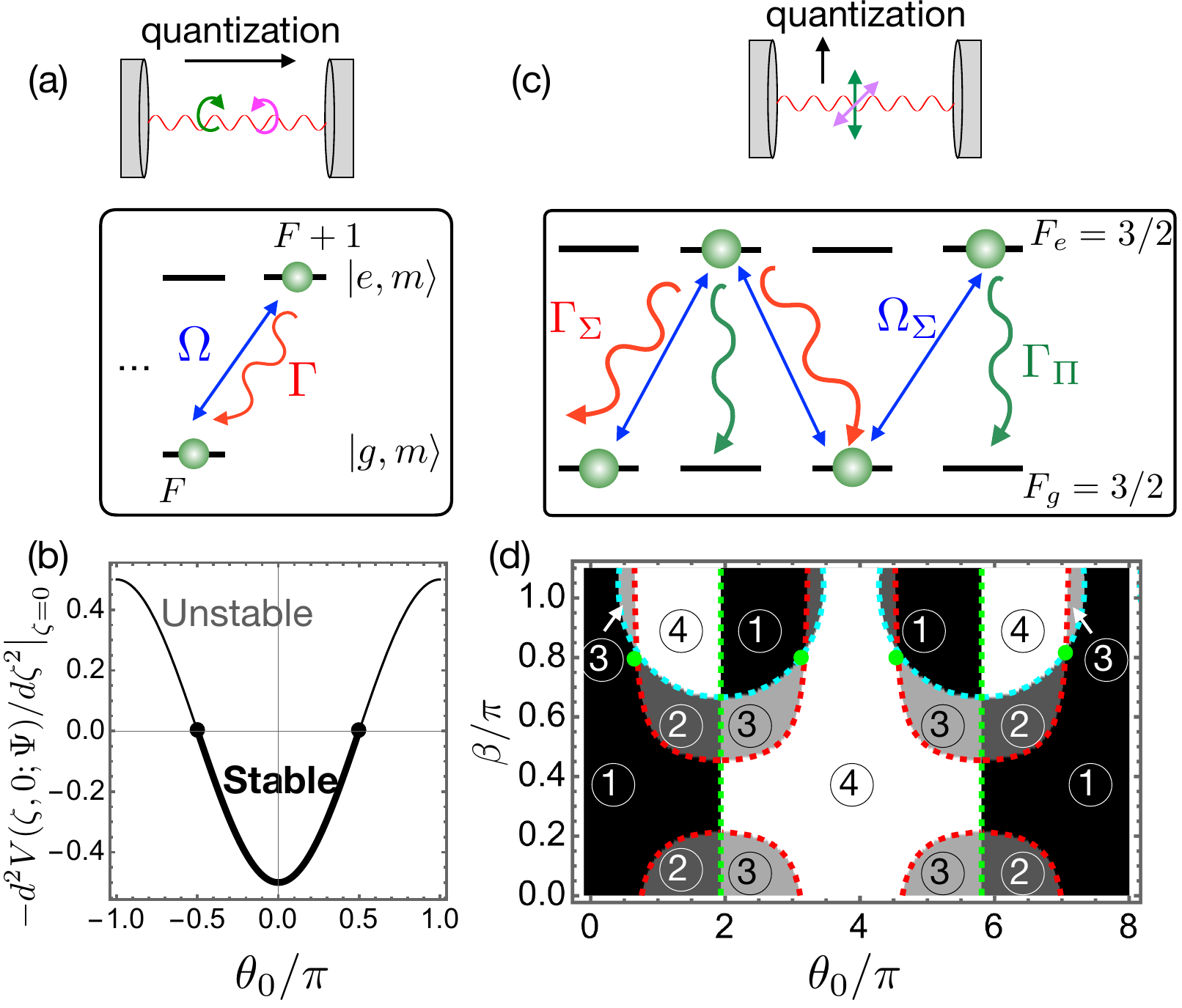}
\caption{(a) An ensemble of effective two-level atoms is driven by a right-circularly polarized laser with strength $\Omega$ and superradiantly decay at rate $\Gamma$ to the ground state. The quantization axis is parallel to the cavity axis. (b) Negative of the curvature of the potential, $-d^2V(\zeta,0;\Psi)/d\zeta^2\big\vert_{\zeta=0}$. There are two critical points, at $\theta_0 = \pm\pi/2$, indicated by two dots. The thick black line marks the stable region. (c) An ensemble of eight-level atoms is driven by a $\Sigma$-polarized laser, and superradiantly emit $\Sigma$- and $\Pi$-polarized light. The quantization axis is perpendicular to the cavity axis. (d) Regions of stability and instability vs the angle $\beta$ in the ground state manifold and state preparation pulse area $\theta_0$ [see text]. The system is stable to emission of both polarizations in regions marked 1, unstable to emission of $\Sigma$-polarized light in region 2, unstable to emission of $\Pi$-polarized light in region 3, and unstable to emission of both in region 4. Green lines and dots mark critical manifolds and points where the system crosses from a stable or unstable region to an unstable or stable region for each polarization, and cyan and red lines show the critical manifolds where emission of only one polarization crosses from stable to unstable.}
\label{fig: potentials}
\end{figure}

We now introduce the concept of a superradiance potential as a visual aid to understand Eq.~\eqref{eqn: stability cond}. For any state $\ket{\Psi}$, we define the potential $V(\zeta_0,\zeta_1; \Psi)$ as
\begin{equation}\label{eqn: superradiance potential}
V(\zeta_0,\zeta_1; \Psi) = \frac{1}{N}\braket{\Psi \vert e^{i(\zeta_0 \hat{D}^x_{\alpha_0} + \zeta_1 \hat{D}^x_{\alpha_1})} \hat n_e e^{-i(\zeta_0 \hat{D}^x_{\alpha_0} + \zeta_1 \hat{D}^x_{\alpha_1})} \vert \Psi},
\end{equation}
where $\zeta_i$ have a similar interpretation to $\theta_0$ in Eq.~\eqref{eqn: Psi}, and $\hat n_e = \sum_i^N \sum_m \ket{e,m}_i\bra{e,m}_i$ is the occupation in the excited states.

The matrix $\mathcal{H}$ in Eq.~\eqref{eqn: stability cond} is proportional to the Hessian matrix of $V$ evaluated at $\zeta_0=\zeta_1 = 0$ and for $\ket{\Psi} = \ket{\Psi(\alpha_0,\theta_0;\vec{\beta})}$:
\begin{equation}
\mathcal{H} = 2N\left(\begin{array}{cc}
\frac{\partial^2V}{\partial\zeta_0^2} & \frac{\partial^2V}{\partial\zeta_0\partial\zeta_1} \\
\frac{\partial^2V}{\partial\zeta_0\partial\zeta_1} & \frac{\partial^2V}{\partial\zeta_1^2}
\end{array}\right)_{\zeta_0=\zeta_1=0}.
\end{equation}
\Rev{ To see this, note that 
the second derivative of $V$ with respect to $\zeta_0$, at $\zeta_0 = \zeta_1 = 0$ is
\begin{align}
\frac{\partial^2}{\partial\zeta_0^2}V(\zeta_0,0; \Psi)\big\vert_{\zeta_0=0} = \frac{1}{N}\braket{\Psi \vert [i\hat D^x_{\alpha_0}, [i\hat D^x_{\alpha_0}, \hat n_e]] \vert \Psi},
\end{align}
which is equal to $\frac{\lambda_{00}}{2N}$. Similarly, $\frac{\partial^2V}{\partial\zeta_1^2}\big\vert_{\zeta_0 = \zeta_1 = 0} = \frac{\lambda_{11}}{2N}$. These give the two diagonal elements of $\mathcal{H}$. In using the product rule to calculate the mixed derivative $\frac{\partial^2V}{\partial\zeta_0\partial\zeta_1}\big\vert_{\zeta_0 = \zeta_1 = 0}$, the expression involves either one derivative on the unitaries on each side of $\hat n_e$ or both derivatives on the same side. Noting that
\begin{align}
& \frac{\partial}{\partial\zeta_0} e^{-i(\zeta_0 \hat{D}^x_{\alpha_0} + \zeta_1 \hat{D}^x_{\alpha_1})}\big\vert_{\zeta_0 = \zeta_1 = 0} = -i\hat{D}^x_{\alpha_0} \\
& \frac{\partial}{\partial\zeta_1} e^{-i(\zeta_{\alpha_0} \hat{D}^x_{\alpha_0} + \zeta_1 \hat{D}^x_{\alpha_1})}\big\vert_{\zeta_0 = \zeta_1 = 0} = -i\hat{D}^x_{\alpha_1} \\
& \frac{\partial^2}{\partial\zeta_0\partial\zeta_1} e^{-i(\zeta_0 \hat{D}^x_{\alpha_0} + \zeta_1 \hat{D}^x_{\alpha_1})}\big\vert_{\zeta_0 = \zeta_1 = 0} = -\frac{ \hat{D}^x_{\alpha_0} \hat{D}^x_{\alpha_1} + \hat{D}^x_{\alpha_1} \hat{D}^x_{\alpha_0}}{2},
\end{align}
we obtain that $\frac{\partial^2V}{\partial\zeta_0\partial\zeta_1}\big\vert_{\zeta_0 = \zeta_1 = 0} = \frac{\lambda_{01}+\lambda_{10}}{4N}$.
}
The stability condition, i.e.,~the requirement that both the eigenvalues of $\mathcal{H}$ are positive, therefore corresponds to the requirement that the potential has positive curvature along all $(\zeta_0,\zeta_1)$ directions at $\zeta_0=\zeta_1 = 0$. This condition generalizes the stability condition for single-polarization potential described in Refs.~\cite{orioli2022emergent, sundar2023squeezing} to consider fluctuations along arbitrary polarizations $(\alpha_0, \alpha_1)$.

As illustrative examples, we plot the eigenvalues of the Hessian matrix for two specific parameter regions of the initial state $\ket{\Psi} = \ket{\Psi(\alpha_0,\theta_0;\vec{\beta})}$ in Fig.~\ref{fig: potentials}. We choose an effective two-level system in Fig.~\ref{fig: potentials}(a-b), and $F_g=F_e=3/2$ in Fig.~\ref{fig: potentials}(c-d) for concreteness. We will calculate the spin squeezing in these examples later in Secs.~\ref{sec: two-level} and~\ref{sec: multilevel}. Our arguments, however, are general and work for any $F_g, F_e$ and $\ket{\Psi(\alpha_0,\theta_0;\vec{\beta})}$.

In the first example [Fig.~\ref{fig: potentials}(a-b)], we consider an effective two-level system ($F_g=F$, $F_e=F+1$), realized with the levels $\ket{g,F}$ and $\ket{e,F+1}$ and we choose the quantization axis as the cavity axis [Fig.~\ref{fig: setup}(b)]. We initialize the atoms in $\ket{g,F}$ and drive the system with right-circularly polarized light, thus preparing $\ket{\Psi(R,\theta_0;\vec{\beta})} = \exp( -i \theta_0 \hat{D}^x_R) \ket{g,F}^{\otimes N}$, with parameters that satisfy Eqs.~\eqref{eqn: mf steady state} and~\eqref{eqn: Omega steady state}. In this case, only the right-handed polarization is relevant, and therefore there is only one nontrivial eigenvalue for $\mathcal{H}$, plotted in Fig.~\ref{fig: potentials}(b). The stable region corresponds to $|\theta_0|\leq \pi/2$ (marked by a thick black line).

In the second example [Fig.~\ref{fig: potentials}(c-d)], we consider an eight-level system with $F_g = F_e = 3/2$ where all levels and both cavity polarizations are relevant. We initialize the atoms in $\ket{G_{\beta}} = \cos\frac{\beta}{2} \ket{g,-\frac{3}{2}} + \sin\frac{\beta}{2} \ket{g,\frac{1}{2}}$. For simplicity, we choose the quantization axis to be perpendicular to the cavity axis [Fig.~\ref{fig: setup}(c)]. We then drive the system with a $\Sigma$-polarized laser: $\ket{\Psi(\Sigma,\theta_0;\beta)} = e^{-i\theta_0\hat{D}^x_\Sigma} \ket{G_{\beta}}^{\otimes N}$. This choice of quantization axis makes $\mathcal{H}$ diagonal \footnote{The superradiant dynamics (in the absence of drive) of this particular eight-level structure with this type of initial condition was also analyzed in Ref.~\cite{orioli2022emergent}. The stability criterion in that work only considered the diagonal of the Hessian, which was correct for the cases considered there. However, as we have shown here, for more general initial conditions the full Hessian needs to be taken into account.}.
Figure~\ref{fig: potentials}(d) shows the regions where the two diagonal elements are positive or negative, with the stable phase (in black) being the one where both are positive.

The superradiance potential has further significance beyond the stability criterion. 
\Rev{To see this, note that the derivative of $V$ with respect to $\zeta_\alpha$ at $\zeta_0 = \zeta_1 = 0$ is
\begin{align}
\frac{\partial V}{\partial\zeta_\alpha} \vert_{\zeta_0=\zeta_1=0} =& \frac{i}{N}\braket{\Psi \vert [\hat D^x_\alpha, \hat n_e] \vert \Psi}\nonumber\\
=& \frac{1}{N}\braket{\Psi \vert \hat D^y_\alpha \vert \Psi}.
\end{align}
The right-hand side in the second line is $\Omega_\alpha/N\Gamma$ in the steady state [Eq.~\eqref{eqn: Omega steady state}], therefore}
\begin{equation}
\Omega_\alpha = N\Gamma\frac{\partial V}{\partial\zeta_\alpha} \big\vert_{\zeta_0=\zeta_1=0}.
\end{equation}
Thus, the slope of $V$ helps determine the location of MF steady states. In the examples above, $\Omega_L \propto dV/d\zeta_L = 0$ in Fig.~\ref{fig: potentials}(c), and $\Omega_\Pi \propto dV/d\zeta_\Pi = 0$ in Fig.~\ref{fig: potentials}(d).
Furthermore, we showed in previous works~\cite{orioli2022emergent, sundar2023squeezing} that when there is only one relevant polarization, the superradiance potential fully describes the mean-field time evolution~\footnote{In Refs.~\cite{orioli2022emergent, sundar2023squeezing}, the convention for defining the one-polarization superradiance potential $V(\zeta)$ was $V(\zeta) \equiv V(\zeta,0;\ket{\Psi(\alpha_0,0;\vec{\beta})})$, where the dependence on $\alpha_0$ and $\vec{\beta}$ was usually suppressed. The derivatives were evaluated at $\zeta = \theta_0$ in \cite{orioli2022emergent, sundar2023squeezing}.}.
However, we note that in general a two-parameter potential $V(\zeta_0,\zeta_1)$ cannot describe the mean-field dynamics when both polarizations are relevant, because of the noncommutativity of $\hat{D}^\pm_{\alpha_0}$ and $\hat{D}^\pm_{\alpha_1}$.

\subsection{Critical manifolds}\label{subsec: critical manifold}
Critical manifolds are manifolds where one or both eigenvalues go to zero, i.e., $\det(\mathcal{H})=0$. Typically, these manifolds separate regions where one or both eigenvalues of $\mathcal{H}$ have opposite signs, i.e., regions that are stable and unstable to either one or both polarizations. 
The critical manifolds will be crucial when we study dissipative squeezing generation, because the system acquires scalable squeezing near the critical regions.

The critical points for the two-level system in Fig.~\ref{fig: potentials}(a) are at $\theta_0 = \pm \frac{\pi}{2}$ (black dots in Fig.~\ref{fig: potentials}(b)). As we will show in Sec.~\ref{sec: two-level}, the system acquires scalable squeezing in one mode near these critical points.

For multilevel systems where two polarizations are relevant, the system can be critical to emission in one polarization, i.e., one of the eigenvalues of $\mathcal{H}$ is zero, indicated by red or blue dashed lines in Fig.~\ref{fig: potentials}(d), or the system can be critical to emission in two polarizations, i.e., $\mathcal{H} = 0$, indicated by green dashed lines or dots. In Fig.~\ref{fig: potentials}(d), there are two lines and four points in the $(\zeta_0,\zeta_1)$ plane where $\mathcal{H} = 0$. 
The system is stable to emission of both polarizations in regions marked 1, unstable to emission of $\Sigma $-polarized light in region 2, unstable to emission of $\Pi$-
polarized light in region 3, and unstable to emission of both
in region 4.
For the multilevel example, we will show that it is possible to generate scalable squeezing in four different quadratures near critical lines between regions 1 and 4, whereas only two squeezed directions can be created close to critical lines between regions 1 and 2 or 1 and 3. 

We emphasize that the only region where the MF state is a good approximation of the full quantum steady state is where the system is stable to both polarizations, i.e.,~the black region in Fig.~\ref{fig: potentials}(d). Outside this region, quantum fluctuations destabilize this MF state and drive it towards a mixture of stable steady states. Calculating the steady state for an initial state in the unstable region is outside the scope of this paper.
In the remainder of this paper, we will focus on the properties of the steady state in the stable region.

It is also reasonable to ask what the critical lines are separating exactly. In the two-level system it is well-known that the critical points are associated to a normal to superradiant phase transition~\cite{walls1978non, drummond1978volterra, drummond1980multiple, drummond1980observables, carmichael1980analytical, puri1979exact, barberena2019driven, wolfe2014spin, somech2022quantum}. The stable region corresponds to the system being in the superradiant phase, where the quantum steady state is close to the MF state. The normal phase corresponds to the case where $\Omega_\alpha/N\Gamma$ is large enough that the system oscillates forever in the MF approximation. In the multilevel system, a similar normal to superradiant phase transition can take place, but other possibilities can emerge as well, such as superradiant to superradiant transitions. This will be investigated in future work.

\section{Quantum correlations}\label{sec: correlations}
Even though we initialize the system in a mean-field steady state, the quantum fluctuations around this state are not stationary. \Rev{They lead to the development of nonzero connected spin correlations, $\braket{\hat S_\mu \hat S_\nu}-\braket{\hat S_\mu}\braket{\hat S_\nu}$, and under certain conditions, quantum entanglement. Here, $\hat S_\mu$ are collective Gellmann spin operators. We will next calculate some of these connected spin correlations, focusing in particular on correlations between spin variables orthogonal to the mean Bloch vector.} We will show that the dissipative quantum dynamics towards the full quantum steady state \Rev{leads to nonzero correlations between these spin variables}, and in our case, to the formation of entanglement between the atoms manifested in the form of spin squeezing~\cite{vitagliano2011spin, vitagliano2014spin, sundar2023squeezing}.
We treat these quantum fluctuations as a small perturbation around the mean-field state via bosonic degrees of freedom in the large-$N$ approximation, which allows us to analytically compute the value of the variances in the bosonic quadratures at the steady state.

This calculation proceeds in four steps, described in detail in Secs.~\ref{subsec: Schwinger bosons}-\ref{subsec: correlations}, and exemplified in Secs.~\ref{sec: two-level} and~\ref{sec: multilevel}. First, we define an exact map between collective spin operators and Schwinger bosons. Second, we make the master equation quadratic in boson operators by making a large-$N$ approximation. Third, we diagonalize the master equation by making a Bogoliubov transformation. And fourth, we solve the master equation.
In this way, we demonstrate the presence of spin squeezing. 
We determine the finite-size scaling of the best achievable squeezing by including higher-order terms in the large-$N$ approximation.

\subsection{Schwinger bosons}\label{subsec: Schwinger bosons}
We can define $\ell$ Schwinger bosons for a collective system of atoms with $\ell$ relevant internal atomic levels, i.e.,~those levels that participate in the dynamics. The most straightforward way to set the Schwinger bosons is by defining bosonic operators $\hat a_{g(e),m}$ which annihilate a particle in $\ket{g(e),m}$. However, this choice is inconvenient because the mean-field state $\ket{\Psi(\alpha_0,\theta_0; \vec{\beta})} = \ket{\psi(\alpha_0,\theta_0; \vec{\beta})}^{\otimes N}$ is in a superposition of states created by $\hat a_{g(e),m}$. A more convenient choice is to define Schwinger boson operators $\hc_\mu$ which annihilate particles in a different orthonormal manifold of states $\ket{\mu},\ \mu\in[0,\ell-1]$, where $\ket{\mu=0}$ is defined as $\ket{0} \equiv \ket{\psi(\alpha_0,\theta_0;
\vec{\beta})}$. We call these Schwinger c-bosons. The basis states $\ket{\mu}$ are related to $\ket{g(e),m}$ by a unitary transformation. Our main results do not depend on this basis choice, but choosing $\ket{0}$ in this way simplifies the calculation. For brevity, we will hereafter drop the symbols $\alpha_0$, $\theta_0$, and $\vec{\beta}$ from $\hat c_\mu(\alpha_0, \theta_0;\vec{\beta})$.

Any collective spin operator can be formally expressed in this basis in a matrix form. For example, we can write the jump operators as $\hat{\mathscr{D}}^-_\alpha = \sum_{i\mu\nu} g_{\alpha,\mu\nu} \ket{\mu}_i \bra{\nu}_i$, where $g_{\alpha,\mu\nu}$ are their matrix elements. In terms of the Schwinger c-bosons, the jump operators then have a quadratic form:
\begin{equation}\label{eqn: gmunu}
\hat{\mathscr{D}}^-_\alpha = \sum_{\mu,\nu} g_{\alpha,\mu \nu}^{\phantom\dagger} \hat c_\mu\+ \hat c_\nu^{\phantom\dagger}.
\end{equation}
Due to the mean-field stationary state conditions, Eqs.~\eqref{eqn: mf steady state} and~\eqref{eqn: Omega steady state}, we have that the coefficient $g_{\alpha,00} = \frac{1}{N}\braket{\Psi(\alpha_0,\theta_0; \vec{\beta}) \vert \hat{\mathscr{D}}^-_\alpha\vert \Psi(\alpha_0,\theta_0; \vec{\beta})} = 0$.

\subsection{The Holstein-Primakoff approximation}\label{subsec: HP}
If the quantum state $\rho$ is close to the mean-field state $\rho_{\rm MF}$, which is a macroscopically occupied state of the $\hc_0$ operator, we can assume that $\rho$ also has macroscopic occupation for $\hc_0$, i.e., $\braket{\hat c_0\+ \hat c_0} \simeq N$ at all times. Therefore, we make the generalized Holstein-Primakoff (HP) approximation $\hc_0 \approx \sqrt{N}$~\cite{kurucz2010multilevel}. Under this approximation, the jump operators simplify to
\begin{align}\label{eqn: Dminus HP}
\hat{\mathscr{D}}^-_\alpha = \sqrt{N} \sum_{\mu>0} (x_{\alpha,\mu}^{\phantom\dagger} \hat X^c_\mu + i y_{\alpha,\mu}^{\phantom\dagger} \hat Y^c_\mu) + \sum_{\mu,\nu>0} g_{\alpha,\mu\nu}^{\phantom\dagger} \hat c_\mu\+ \hat c_\nu^{\phantom\dagger},
\end{align}
where $\hat X^c_\mu = \frac{\hat c_\mu^{\phantom\dagger} + \hat c_\mu\+}{\sqrt{2}}$ and $\hat Y^c_\mu = \frac{\hat c_\mu^{\phantom\dagger} - \hat c_\mu\+}{i\sqrt{2}}$ are the real and imaginary parts of $\hat c_\mu$ and are analogous to position and momentum quadratures. The coefficients $x_{\alpha,\mu}$ and $y_{\alpha,\mu}$ are given by \begin{align}
& x_{\alpha,\mu} + y_{\alpha,\mu} = \sqrt{2}\braket{0 \vert_j^{\phantom\dagger}\ \hat d^-_{j,\alpha} \vert \mu}_j, \nonumber\\
& x_{\alpha,\mu} - y_{\alpha,\mu} = \sqrt{2}\braket{\mu \vert_j^{\phantom\dagger}\ \hat d^-_{j,\alpha} \vert 0}_j.
\end{align}

As we will show, the $x_{\alpha,\mu}$ and $y_{\alpha,\mu}$ terms in $\hat{\mathscr{D}}^-_\alpha$ determine the leading-order $[O(1)]$ behavior of the quantum correlations, while the $g_{\alpha,\mu\nu>0}$ terms lead to finite-size corrections of order $O(1/N)$. For brevity, we will collect the components of $x_{\alpha,\mu}$ and $y_{\alpha,\mu}$ into the vectors $\vec{x}_\alpha$ and $\vec{y}_\alpha$, and the components $g_{\alpha,\mu\nu>0}$ into the matrix $\overleftrightarrow{g}_\alpha$.

While the values of $x_{\alpha,\mu}$ and $y_{\alpha,\mu}$ are related to matrix elements of $\hat{\mathscr{D}}^-_\alpha$, and are therefore basis dependent, physically relevant quantities such as the curvature of the superradiance potential, critical points, and spin squeezing do not depend on the basis choice. Instead, they only depend on the physical parameters $(\alpha_0,\theta_0; \vec{\beta})$. For example, to leading order the Hessian matrix $\mathcal{H}$ can be written as
\begin{align}
&\mathcal{H} = \nonumber\\ & N \left(\begin{array}{cc}
2{\rm Re}(\vec{x}^*_{\alpha_0} \cdot \vec{y}_{\alpha_0}) &
{\rm Re}(\vec{x}^*_{\alpha_0} \cdot \vec{y}_{\alpha_1} + \vec{x}^*_{\alpha_1} \cdot \vec{y}_{\alpha_0})
\\
{\rm Re}(\vec{x}^*_{\alpha_0} \cdot \vec{y}_{\alpha_1} + \vec{x}^*_{\alpha_1} \cdot \vec{y}_{\alpha_0})
&
2{\rm Re}(\vec{x}^*_{\alpha_1} \cdot \vec{y}_{\alpha_1})
\end{array}\right).
\end{align}
Basis rotations lead to SU($\ell-1$) rotations of $\vec{x}_\alpha$ and $\vec{y}_\alpha$, and their dot products are invariant under SU($\ell-1$) rotations.
For our examples, $x_{\alpha,\mu}$ and $y_{\alpha,\mu}$ are real, and as such we will set them to be real hereafter.

\subsection{Bogoliubov transformation}\label{subsec: Bogoliubov}
To leading order in $N$, i.e.,~ignoring $g_{\alpha,\mu\nu>0}$ in Eq.~(\ref{eqn: Dminus HP}), the jump operators are linear and the master equation is quadratic in the Schwinger bosonic variables in the HP approximation. Thus, we can analytically solve the system using a Bogoliubov transformation.
For this purpose, note first that at this order, the jump operator $\hat{\mathscr{D}}^-_\alpha \simeq \sqrt{N}\sum_{\mu>0} (x_{\alpha,\mu}^{\phantom\dagger} \hat X^c_\mu + i y_{\alpha,\mu}^{\phantom\dagger} \hat Y^c_\mu)$ can be interpreted as being proportional to a single annihilation operator. Specifically, we define two Bogoliubov operators $\hat b_{\alpha_0}$ and $\hat b_{\alpha_1}$ as
\begin{equation}\label{eqn: bogoliubons for D-}
\hat b_\alpha \equiv \frac{1}{\sqrt{2N \vec{x}_\alpha \cdot \vec{y}_\alpha }} \hat{\mathscr{D}}^-_\alpha,\quad \alpha \in \{\alpha_0, \alpha_1\}.
\end{equation}
We call them Bogoliubov b-bosons.
Importantly, the steady-state condition $\hat{\mathscr{D}}^-_\alpha\rho = 0$ implies that the steady state is the vacuum of
$\hat b_{\alpha_0}$ and $\hat b_{\alpha_1}$.

The commutators $[\hat b_\alpha^{\phantom\dagger}, \hat b_{\alpha'}\+]$ are proportional to the Hessian matrix elements in Eq.~\eqref{eqn: stability cond}.
The normalization factor in Eq.~\eqref{eqn: bogoliubons for D-} ensures that $[\hat b_{\alpha}^{\phantom\dagger}, \hat b_{\alpha}\+] = 1$. If $\mathcal{H}$ is not diagonal, then $[\hat b_{\alpha_0}, \hat b_{\alpha_1}\+]$ is nonzero. In this case, a convenient method is to first find the basis of atomic jump operators that diagonalizes $\mathcal{H}$, i.e.,~a convenient polarization basis, and then define Bogoliubov operators corresponding to those jump operators. Thus, without loss of generality, we can assume that $[\hat b_{\alpha_0}, \hat b_{\alpha_1}\+]=0$ and the Hessian is diagonal.

Since there are $(\ell-1)$ Schwinger c-bosons $\hc_\mu$ with $\mu>0$, there have to be $(\ell-3)$ other independent Bogoliubov b-bosons, $\hat b_\gamma, \gamma\in[1,\ell-3]$, in addition to $\hat b_{\alpha_0}$ and $\hat b_{\alpha_1}$. These Bogoliubov operators commute with $\hat{\mathscr{D}}^\pm_\alpha$, and correspondingly are conserved during the evolution to the steady state in this approximation. Despite their dynamics being trivial, they can still play an important role in shaping the dynamics of the Schwinger c-bosons, as will be explained in Sec.~\ref{sec: multilevel}.

\subsection{Calculating the quantum correlations}\label{subsec: correlations}
Starting from $\ket{\Psi(\alpha_0,\theta_0;\vec{\beta})}$, which is the vacuum of $\hat c_{\mu > 0}$, the driven-dissipative dynamics leads to the development of correlations between the bosonic fields $\hat c_{\mu > 0}$. \Rev{ The bosonic variables $\hat X^c_\mu$ and $\hat Y^c_\mu$ approximate the collective spin variables $\hat S^x_\mu = \sum_j (\ket{0}\bra{\mu} + {\rm h.c.})/\sqrt{8N}$ and $\hat S^y_\mu = \sum_j (i\ket{0}\bra{\mu} + {\rm h.c.})/\sqrt{8N}$. Therefore, the quantum correlations between the bosonic variables, e.g. $\braket{\hat X^c_\mu\hat X^c_\nu}-\braket{\hat X^c_\mu}\braket{\hat X^c_\nu}$ and $\braket{\hat Y^c_\mu\hat Y^c_\nu}-\braket{\hat Y^c_\mu}\braket{\hat Y^c_\nu}$, capture correlations between the spin variables.
}

We quantify the correlations via the covariance matrix $\Xi^c = \left( \begin{array}{cc} \Xi^c_{XX} & \Xi^c_{XY} \\ (\Xi^c_{XY})^T & \Xi^c_{YY} \end{array}\right)$, where $\Xi^c_{XX}, \Xi^c_{XY}$, and $\Xi^c_{YY}$ are the covariance matrices for the Schwinger c-boson variables $\hat X^c$ and $\hat Y^c$:
\begin{align}\label{eqn: Xic defn}
& (\Xi^c_{XX})_{\mu\nu} = \braket{ \hat X^c_\mu \hat X^c_\nu + \hat X^c_\nu \hat X^c_\mu} - 2\braket{ \hat X^c_\mu} \braket{\hat X^c_\nu}, \nonumber\\
& (\Xi^c_{XY})_{\mu\nu} = \braket{\hat X^c_\mu \hat Y^c_\nu + \hat Y^c_\nu \hat X^c_\mu} - 2\braket{\hat X^c_\mu} \braket{\hat Y^c_\nu}, \nonumber\\
& (\Xi^c_{YY})_{\mu\nu} = \braket{\hat Y^c_\mu \hat Y^c_\nu + \hat Y^c_\nu \hat Y^c_\mu} - 2 \braket{ \hat Y^c_\mu} \braket{ \hat Y^c_\nu},
\end{align}
and $\mu,\nu>0$.
At $t=0$, $\Xi^c$ is the identity matrix.

In the HP approximation, the bosonic operators $\hc_\mu$ approximate the spin operators $\hat\Lambda_\mu = \frac{1}{\sqrt{N}}\sum_i \ket{0}_i\bra{\mu}_i$, which are $2(\ell-1)$ spin variables perpendicular to the collective spin vector. Therefore, the matrix elements of $\Xi^c$ are approximately equal to covariances of the real and imaginary parts of $\hat\Lambda_\mu$. These are the only relevant variables as they have $O(1)$ fluctuations in the initial state $\ket{0}$; the remaining orthogonal variables $\hat\Lambda_{\mu\nu} = \frac{1}{\sqrt{N}}\sum_i \ket{\mu}_i\bra{\nu}_i$ with $\mu,\nu>0$ are suppressed by $1/\sqrt{N}$. Therefore, the covariance matrix $\Xi^c$ describes the quantum noise (normalized spin variances) perpendicular to the collective spin vector. Any eigenvalue of $\Xi^c$ decreasing below 1 indicates a reduction in spin projection noise perpendicular to the collective spin vector, as compared to the initial coherent state. Such noise reduction perpendicular to the collective spin vector is spin squeezing in a multilevel system when the spin length is order $N$~\cite{kurucz2010multilevel, vitagliano2011spin}, and is analogous to spin-squeezing for spin-1/2 atoms.

The simplest way to calculate $\Xi^c$ is in the Bogoliubov framework. As argued above, the dynamics brings the system to a steady state where the occupation of the Bogoliubov modes associated to $\hb_{\alpha_0}$ and $\hb_{\alpha_1}$ relaxes to the vacuum value, any correlations associated with $\hb_{\alpha_0}$ or $\hb_{\alpha_1}$ decay to zero, and correlations of all other Bogoliubov modes are left untouched. Analogous to $\Xi^c$, we define $\Xi^b = \left( \begin{array}{cc} \Xi^b_{XX} & \Xi^b_{XY} \\ (\Xi^b_{XY})^T & \Xi^b_{YY} \end{array}\right)$ where $\hat X^b_\mu = \frac{\hb_\mu^{\phantom\dagger} + \hb_\mu\+}{\sqrt{2}}$ and $\hat Y^b_\mu = \frac{\hb_\mu^{\phantom\dagger} - \hb_\mu\+}{i\sqrt{2}}$, and $\Xi^b_{XX}$, $\Xi^b_{YY}$, and $\Xi^b_{XY}$ are defined similar to Eq.~\eqref{eqn: Xic defn} but in terms of quadratures of Bogoliubov b-bosonic operators instead of Schwinger c-bosons. We obtain $\Xi^c$ by inverting the Bogoliubov transformation, and the squeezing can be inferred from the eigenvalues of $\Xi^c$. Note that since the Bogoliubov transformation is not unitary, the eigenvalues of $\Xi^c$ are different from those of $\Xi^b$.

In the following sections, we show concrete examples of this procedure using effective two-level [Sec.~\ref{sec: two-level}] and multilevel [Sec.~\ref{sec: multilevel}] systems.

\section{Two-level system}\label{sec: two-level}
\begin{figure}[t]
\includegraphics[width=1.0\columnwidth]{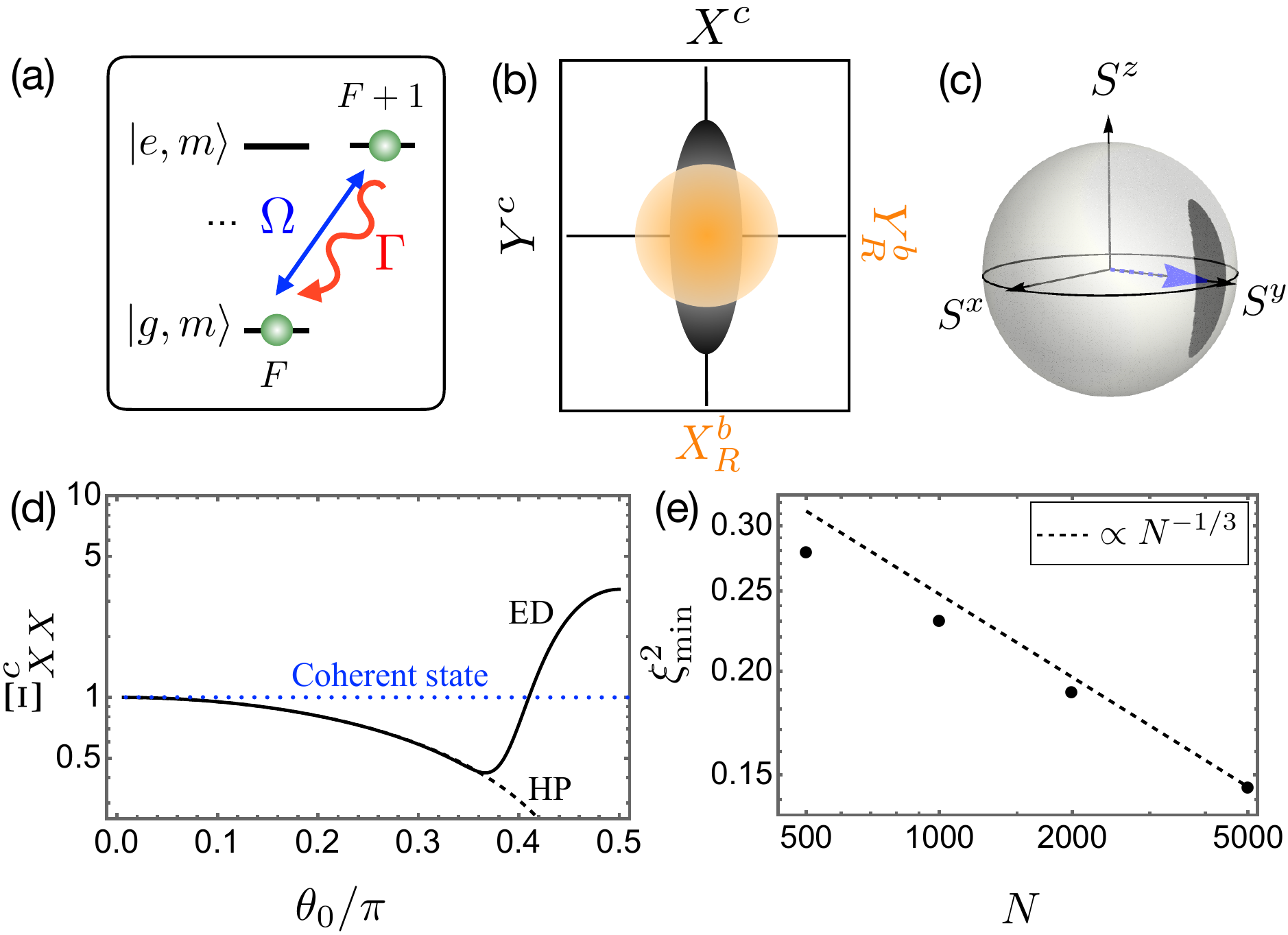}
\caption{(a) An ensemble of effective two-level atoms [see also Fig.~\ref{fig: potentials}(a)]. (b) Illustration of the steady-state squeezing in bosonic quadratures. The steady state is the coherent vacuum of $\hat X^b$ and $\hat Y^b$, which makes it squeezed in $\hat X^c$ and antisqueezed in $\hat Y^c$ [see text for definitions of the quadratures]. (c) Visualizing the collective spin squeezing on a Bloch sphere in the two-level system. The squeezing is along $\hat S^x$. The visualization shows a particular example where the steady state is near $\theta_c = \pi/2$. (d) Steady-state squeezing vs $\theta_0$ in the two-level case, obtained from an exact numerical calculation with $N=100$ atoms (solid line). The black dashed line plots the HP prediction, $\cos\theta_0$, and the blue dotted line plots $\Xi^c_{XX}$ in the coherent state. (e) The best squeezing achievable vs $N$ has a scaling close to $N^{-1/3}$.}
\label{fig: 2 level}
\end{figure}

First, we review the driven-dissipative dynamics of an effective two-level system realized within the $\ket{g,F}$ and $\ket{e,F+1}$ manifold of a system with $F_g=F$ and $F_e = F+1$, when driven by right-circularly polarized light as shown in Figs.~\ref{fig: potentials}(a) and~\ref{fig: 2 level}(a). Even though the driven-dissipative dynamics of two-level systems has been studied extensively in the literature~\cite{walls1978non, drummond1978volterra, drummond1980multiple, drummond1980observables, carmichael1980analytical, puri1979exact, barberena2019driven, wolfe2014spin, somech2022quantum} we discuss it first to facilitate the understanding of the more complex multilevel systems presented below.

The jump operator for emission of right-circularly polarized light is $\hat{\mathscr{D}}^-_R$. The left-handed polarization is irrelevant. The only relevant term in $\hat{\mathscr{D}}^-_R$ for the two levels is $C^F_R \hat S^-_{F,R}$ with $\hat S^-_{F,R} = \sum_i \hat s^-_{F,i,R}$, and all the relevant dynamics can be visualized on one Bloch sphere whose axes are $\hat S^\alpha_{F,R}, \alpha\in\{x,y,z\}$. The Clebsch-Gordan coefficient for this transition is $C^F_R = 1$. For brevity in this section, we will refer to $\hat S^\alpha_{F,R}$ as simply $\hat S^\alpha$, dropping the subscripts referring to the angular momentum and polarization.
The mean spin direction on this Bloch sphere, i.e., the mean Bloch vector, initially points along $\hat{S}_{\rm Bloch} = (0,\sin\theta_0, - \cos\theta_0)$, where $\theta_0$ is the angle that the Bloch vector makes with the south pole.

From Sec.~\ref{sec: mf}, the superradiance potential for the two-level system is $V(\zeta,0;\Psi) = \sin^2\frac{\zeta+\theta_0}{2}$. The mean-field state is stationary if $\Omega_R = \frac{N\Gamma}{2}\sin\theta_0$. Figure~\ref{fig: potentials}(b), which plots $d^2V/d\zeta^2\big\vert_{\zeta=0}$ shows that the system is stable in the region $-\frac{\pi}{2} < \theta_0 < \frac{\pi}{2}$, as discussed in Sec.~\ref{sec: mf}.

\subsection{Quantum correlations}\label{subsec: two-level correlations}
The system has quantum fluctuations along the directions $(1,0,0)$ and $(0,\cos\theta_0, \sin\theta_0)$, which are the two directions perpendicular to the Bloch vector $\hat{S}_{\rm Bloch}$ on the Bloch sphere. The driven-dissipative dynamics squeezes and antisqueezes the quantum noise in these orthogonal directions, which we calculate with the Bogoliubov framework in the HP approximation.

To do this, we define two Schwinger c-bosons via
\begin{align}
\hc_0 = \cos\frac{\theta_0}{2}\ha_{g,F} +
i\sin\frac{\theta_0}{2}\ha_{e,F+1} \nonumber\\
\hc_1 = i\sin\frac{\theta_0}{2}\ha_{g,F} + \cos\frac{\theta_0}{2}\ha_{e,F+1}
\end{align}
This determines $g_{\alpha,\mu\nu}$:
\begin{equation}
\overleftrightarrow{g}_R = \left(\begin{array}{cccc}
0 & \frac{\sqrt{N}}{2}(1+\cos\theta_0)\\
\frac{\sqrt{N}}{2}(1-\cos\theta_0) & -\frac{i}{2}\sin\theta_0
\end{array}\right),
\end{equation}
and thus $\hat{\mathscr{D}}^-_R = \sqrt{\frac{N}{2}}(\hat X^c_1 + i\cos\theta_0 \hat Y^c_1) - \frac{i\sin\theta_0}{2}(\hat X^c_1+i\hat Y^c_1)(\hat X^c_1-i\hat Y^c_1)$. The Schwinger c-bosonic variables $\hat X^c_1$ and $\hat Y^c_1$ are proportional to the orthogonal spin variables $\hat S^x$ and $\cos\theta_0 \hat S^y + \sin\theta_0 \hat S^z$, respectively. 
Following Sec.~\ref{subsec: Bogoliubov}, we define the Bogoliubov b-boson $\hat b_R = \frac{\hat X^c_1 + i\cos\theta_0\hat Y^c_1}{\sqrt{2\cos\theta_0}}$. At leading order, $\hat{\mathscr{D}}^-_R = \sqrt{N\cos\theta_0}\,\hat b_R$.

The master equations for the elements of $\Xi^b$ are
\begin{align}\label{eqn: 2 level master eqn}
\partial_t \Xi^b_{XX} = &N\Gamma\cos\theta_0(1 - \Xi^b_{XX}) \nonumber\\ &+ \underbrace{\Gamma\sin^2\theta_0 \left( \frac{\Xi^b_{YY}}{\cos^2\theta_0} - \Xi^b_{XX} \right)}_{\rm finite\ size}, \nonumber\\
\partial_t \Xi^b_{YY} = &N\Gamma\cos\theta_0(1 - \Xi^b_{YY}) \nonumber\\&+ \underbrace{\Gamma\sin^2\theta_0 \left( \Xi^b_{XX}\cos^2\theta_0 - \Xi^b_{YY} \right)}_{\rm finite\ size},\nonumber\\
\partial_t\Xi^b_{XY} = &-(N\Gamma\cos\theta_0+4\Gamma\sin^2\theta_0) \Xi^b_{XY}.
\end{align}
The higher-order terms are highlighted with an underbrace for clarity, and we will use them to calculate the finite-size corrections for the steady-state squeezing. 
Solving Eqs.~\eqref{eqn: 2 level master eqn} and inverting the Bogoliubov transform, we obtain the leading [$O(N\Gamma)$] and subleading [$O(\Gamma)$] terms for the time evolution of the covariance matrix for the Schwinger c-bosons. The solution for $\Xi^c$ due to only the leading $[O(N\Gamma)]$ terms is
\begin{align}\label{eqn: 2 level squeezing}
& \Xi^c_{XX} = \cos\theta_0 + (1-\cos\theta_0)e^{-N\Gamma t\cos\theta_0}, \nonumber\\
& \Xi^c_{YY} = \frac{1}{\cos\theta_0} + \left(1-\frac{1}{\cos\theta_0}\right)e^{-N\Gamma t\cos\theta_0}, \nonumber\\
& \Xi^c_{XY} = 0.
\end{align}
Therefore, $\Xi^c$ is diagonal, and its eigenvalues are $\Xi^c_{XX}$ and $\Xi^c_{YY}$. Of these, $\Xi^c_{XX} < 1$ (for $0<|\theta_0|<\frac{\pi}{2}$), and is therefore squeezed. The squeezing is along $\hat X^c_1 \propto \hat S^x$, and reaches a steady-state value of $\cos\theta_0$ as $t\rightarrow\infty$, at the rate $1/(N\Gamma\cos\theta_0)$. The antisqueezing is along $\hat Y^c_1 \propto \sin\theta_0 \hat S^z + \cos\theta_0 \hat S^y$, with a steady-state value $1/\cos\theta_0$. The squeezing $\Xi^c_{XX}$ approaches zero in the steady state at the critical points $\theta_c = \pm\frac{\pi}{2}$. In Figs.~\ref{fig: 2 level}(b-c) we illustrate the steady-state noise distribution in the Bogoliubov basis ($\hat X^b_R, \hat Y^b_R$), in the Schwinger basis ($\hat X^c_1, \hat Y^c_1$), and on the Bloch sphere.

\subsection{Finite-size corrections in the steady state}
Although to leading order the squeezing at the critical points goes to zero, in reality, higher-order corrections limit the amount of attainable squeezing.

\Rev{The squeezing including higher-order effects can be obtained by solving Eq.~\eqref{eqn: 2 level master eqn}:
\begin{equation}
\left( \begin{array}{c} \Xi^b_{XX} \\ \Xi^b_{YY} \end{array}\right) = e^{-\Lambda t}\left( \begin{array}{c} {\rm sec}\ \theta_0 \\ \cos\theta_0 \end{array}\right) + (1-e^{-\Lambda t})\Lambda^{-1}
\left( \begin{array}{c} N\Gamma\cos\theta_0 \\ N\Gamma\cos\theta_0 \end{array}\right)
\end{equation}
where
\begin{equation}
\Lambda = \left( \begin{array}{cc}
-N\Gamma\cos\theta_0 - \Gamma\sin^2\theta_0 & \Gamma\tan^2\theta_0\\
\Gamma\sin^2\theta_0\cos^2\theta_0 & -N\Gamma\cos\theta_0 - \Gamma\sin^2\theta_0
\end{array}\right).
\end{equation}
Explicit calculation of $e^{-\Lambda t}$ yields
\begin{widetext}
\begin{align}
\Xi^b_{XX} = &\frac{\sec^2\theta_0}{2(N\cos\theta_0+2\sin^2\theta_0)}
\bigg( -N\cos\theta_0\sin^2\theta_0 e^{-\Gamma t(N\cos\theta_0+2\sin^2\theta_0)} 
- e^{-N\Gamma t\cos\theta_0}4\sin^2\frac{\theta_0}{2}(N\cos\theta_0+2\sin^2\theta_0) 
\nonumber\\ & 
+ 2(N\cos^3\theta_0+\sin^2\theta_0(1+\cos^2\theta_0)) \bigg) \nonumber\\
\Xi^b_{YY} = &\frac{1}{2(N\cos\theta_0+2\sin^2\theta_0)} 
\bigg( -N\cos\theta_0\sin^2\theta_0 e^{-\Gamma t(N\cos\theta_0+2\sin^2\theta_0)} 
- e^{-N\Gamma t\cos\theta_0}4\sin^2\frac{\theta_0}{2}(N\cos\theta_0+2\sin^2\theta_0) 
\nonumber\\ & 
+ 2(N\cos\theta_0+\sin^2\theta_0(1+\cos^2\theta_0)) \bigg).
\end{align}
\end{widetext}
The squeezing is $\Xi^c_{XX} = \Xi^b_{XX} \cos\theta_0$, and the antisqueezing is $\Xi^c_{YY} = \Xi^b_{YY}/\cos\theta_0$.

Close to the critical points we have $\cos\theta_0 \approx 0$. Setting $t\rightarrow\infty$ for the steady state and expanding in powers of $\cos\theta_0$, the steady-state squeezing is
}
\begin{equation}
\Xi^c_{XX} \approx \cos\theta_0 + \underbrace{\frac{1}{N\cos^2\theta_0}}_{\rm finite\ size}.
\end{equation}
This shows that the squeezing reaches an optimum value of $\frac{3}{(4N)^{1/3}}$ when the Bloch vector's angle with the south pole is $\theta_0 \sim \frac{\pi}{2} - \left(\frac{2}{N}\right)^{1/3}$. 

Figure~\ref{fig: 2 level}(d) shows the steady-state squeezing vs $\theta_0$, obtained from an exact numerical calculation with $N=100$ atoms. The squeezing agrees well with the HP leading-order prediction $\Xi^c_{XX} = \cos\theta_0$, until finite-size effects kick in and set a limit on squeezing. Figure~\ref{fig: 2 level}(e) shows that the best squeezing reaches an $N^{-1/3}$ scaling in agreement with previous literature~\cite{barberena2019driven, shahmoon2017cooperative, wolfe2014spin, somech2022quantum}, and close to the scaling predicted by our analysis.

\section{Multilevel system}\label{sec: multilevel}

Next, we consider the squeezing generated in multilevel atoms.
In principle, there are multiple level structures and initial conditions one may consider. However, our main conclusions will be the same for most other internal structures or initial states, and are the following.
\begin{enumerate}
\item The system generally hosts two squeezed modes for each relevant cavity polarization. If only one polarization is relevant~\cite{sundar2023squeezing}, two modes are squeezed; if both polarizations are relevant, four modes are squeezed, as we show below.
\item The best squeezing attainable close to the critical point generally scales as $N^{-1/4}$.
\end{enumerate}
We note that there are some fringe cases where the system behaves like a two-level system for emission of one of the polarizations, and the number of squeezed modes is reduced to either 1 or 3. We will explain these fringe cases in Sec.~\ref{subsec: multilevel squeezing}.

To illustrate these findings, we choose the example of Fig.~\ref{fig: potentials}(c) with $F_g = F_e = 3/2$, where all $\ell=8$ internal levels and both cavity polarizations are relevant. We choose to decompose the polarizations in the linear basis, such that the system's evolution is governed by $\hbar\frac{d\rho}{dt} = \mathcal{L}_\Sigma[\rho] + \mathcal{L}_\Pi[\rho]$, where the respective jump operators for $\mathcal{L}_\Sigma$ and $\mathcal{L}_\Pi$ are $\hat{\mathscr{D}}^-_\Sigma$ and $\hat{\mathscr{D}}^-_\Pi$.
 
\subsection{Holstein-Primakoff approximation and Bogoliubov transformation}
The jump operators $\hat{\mathscr{D}}^-_\Sigma$ and $\hat{\mathscr{D}}^-_\Pi$ expressed in terms of the Schwinger c-boson operators are
\begin{align}\label{eqn: multilevel jump operators}
& \hat{\mathscr{D}}^-_\Sigma = \sqrt{N}\sum_{\mu>0} \left(x_{\Sigma,\mu}^{\phantom\dagger} \hat X^c_\mu + i y_{\Sigma,\mu}^{\phantom\dagger} \hat Y^c_\mu\right) + \sum_{\mu\nu>0} g_{\Sigma,\mu\nu}^{\phantom\dagger}\hc_\mu\+ \hc_\nu^{\phantom\dagger}, \nonumber\\
& \hat{\mathscr{D}}^-_\Pi = \sqrt{N}\sum_{\mu>0} \left(x_{\Pi,\mu}^{\phantom\dagger} \hat X^c_\mu + i y_{\Pi,\mu}^{\phantom\dagger} \hat Y^c_\mu\right) + \sum_{\mu\nu>0} g_{\Pi,\mu\nu}^{\phantom\dagger}\hc_\mu\+ \hc_\nu^{\phantom\dagger}.
\end{align}
The values of $x_{\Sigma(\Pi),\mu}$, $y_{\Sigma(\Pi),\mu}$, and $g_{\Sigma(\Pi),\mu\nu}$ depend on the basis states $\ket{\mu}$ used to define the Schwinger c-bosons.  Appendix~\ref{app: Schwinger bosons} \Rev{gives the values of $\vec{x}_{\Sigma(\Pi)}$ and $\vec{y}_{\Sigma(\Pi)}$ for one particular choice of basis:
\begin{widetext}
\begin{equation}\label{eqn: schwinger bosons defn}
\left( \begin{array}{c}
\hc_0 \\ \hc_1 \\ \hc_3 \\ \vdots \\ \hc_7
\end{array}\right)
= 
\left( \begin{array}{cccccccc}
\cos\frac{\beta}{2} & 0 & \sin\frac{\beta}{2} & 0 & \cdots & && \\ 0 & i & 0 & \cdots &&&& \\ -i\sin\frac{\beta}{2} & 0 & i\cos\frac{\beta}{2} & 0 & \cdots & && \\ 0&0&0&i&0&\cdots&& \\ \vdots &&&& 1&\cdots&& \\ \vdots &&&& &\ddots&&
\end{array}\right)\times e^{i\theta_0 d^x_\Sigma}
\left( \begin{array}{c}
\ha_{g,-3/2} \\ \ha_{g,-1/2} \\ \ha_{g,1/2} \\ \vdots \\ \ha_{e,3/2}
\end{array}\right)
\end{equation}
\end{widetext}
where $d^x_\Sigma$ is the matrix for $\hat{d}^x_{i,\Sigma}$ in the $\{\ket{g,m}, \ket{e,m}\}$ basis.}

Following Sec.~\ref{subsec: Bogoliubov}, we define two Bogoliubov b-bosons:
\begin{align}
\hat b_\Sigma = \sum_{\mu=1}^{7} \frac{x_{\Sigma,\mu}^{\phantom\dagger} \hat X^c_\mu + iy_{\Sigma,\mu}^{\phantom\dagger}\hat Y^c_\mu}{\sqrt{2\vec{x}_\Sigma\cdot\vec{y}_\Sigma}}, \nonumber\\
\hat b_\Pi = \sum_{\mu=1}^{7} \frac{x_{\Pi,\mu}^{\phantom\dagger} \hat X^c_\mu + iy_{\Pi,\mu}^{\phantom\dagger}\hat Y^c_\mu}{\sqrt{2\vec{x}_\Pi\cdot\vec{y}_\Pi}},
\label{eq:8l_BogPiSigma_def}
\end{align}
such that $\hat{\mathscr{D}}^-_\Sigma = \sqrt{N\vec{x}_\Sigma\cdot\vec{y}_\Sigma}\, \hat b_\Sigma + O(1)$ and $\hat{\mathscr{D}}^-_\Pi = \sqrt{N\vec{x}_\Pi\cdot\vec{y}_\Pi}\, \hat b_\Pi + O(1)$. Additionally, there are $(\ell-3)=5$ more Bogoliubov b-bosons, which we can write as 
\begin{equation}
\hat b_\nu = \sum_{\mu=1}^7 \frac{y_{\nu,\mu}^{\phantom\dagger} \hat X^c_\mu + ix_{\nu,\mu}^{\phantom\dagger}\hat Y^c_\mu}{\sqrt{2}}.
\label{eq:8l_Bogconserved_def}
\end{equation}
The normalization condition $[\hat b_\nu, \hat b_\nu\+] = 1$ is equivalent to setting $\vec{x}_\nu \cdot \vec{y}_\nu = 1$. Because of the commutation relations and using an appropriate choice of basis, all the $\vec{x}$ vectors can be made mutually orthogonal to each other, and the $\vec{y}$ vectors can be made mutually orthogonal to each other (see Appendix~\ref{app: xy relations}).
This is why for convenience we reversed the definition of $x$ and $y$ in Eq.~(\ref{eq:8l_Bogconserved_def}) compared to Eq.~(\ref{eq:8l_BogPiSigma_def}).

Since $\mathcal{H}$ is diagonal for this choice of polarization basis and $\ket{\Psi(\alpha_0,\theta_0;
\vec{\beta})}$, the system is critical to emission of $\alpha$-polarized light if $\vec{x}_\alpha \cdot \vec{y}_\alpha = 0$, $(\alpha = \Sigma,\Pi)$.

\subsection{Quantum correlations}
Next, we calculate the $14\times14$ covariance matrix for the Bogoliubov b-bosons, and invert the Bogoliubov transformation to get the covariance matrix for the Schwinger c-bosons. As in the two-level system [Sec.~\ref{sec: two-level}], we again have $\Xi^b_{XY} = 0$ at all times for our choice of initial conditions, and therefore $\Xi^b$ is block diagonal. Similar to the two-level system [Eq.~\eqref{eqn: 2 level master eqn}], we write and solve the master equations for the elements of $\Xi^b_{XX}$ and $\Xi^b_{YY}$. 

The initial values for the elements of $\Xi^b_{XX}$ are
\begin{align}
& \braket{\hat X^b_\alpha\hat X^b_\beta} = \frac{\vec{x}_\alpha \cdot \vec{x}_\beta}{4\sqrt{(\vec{x}_\alpha\cdot\vec{y}_\alpha)(\vec{x}_\beta\cdot\vec{y}_\beta)}}, \nonumber\\
& \braket{\hat X^b_\alpha\hat X^b_\mu} = \frac{\vec{x}_\alpha \cdot \vec{y}_\mu}{2\sqrt{2\vec{x}_\alpha\cdot\vec{y}_\alpha}}, \nonumber\\
& \braket{\hat X^b_\mu\hat X^b_\nu} = \frac{1}{2}\vec{y}_\mu \cdot \vec{y}_\nu = \frac{1}{2}\delta_{\mu\nu},
\end{align}
where $\alpha,\beta \in \{\Sigma,\Pi\}$ and $\mu,\nu\neq\Sigma,\Pi$. Their subsequent evolution at leading order is governed by the master equations
\begin{align}\label{eqn: bogoliubov dynamics}
& \partial_t \braket{(\hat X^b_\alpha)^2} = N\Gamma\vec{x}_\alpha\cdot\vec{y}_\alpha (1 - 2\braket{(\hat X^b_\alpha)^2}), \nonumber\\
& \partial_t \braket{\hat X^b_\Sigma \hat X^b_\Pi} = -N\Gamma(\vec{x}_\Sigma\cdot\vec{y}_\Sigma + \vec{x}_\Pi\cdot\vec{y}_\Pi) \braket{\hat X^b_\Sigma \hat X^b_\Pi}, \nonumber\\
& \partial_t \braket{\hat X^b_\alpha \hat X^b_\mu} = -N\Gamma\vec{x}_\alpha\cdot\vec{y}_\alpha \braket{\hat X^b_\alpha \hat X^b_\mu}, \nonumber\\
& \partial_t \braket{\hat X^b_\mu \hat X^b_\nu} = 0.
\end{align}
The solution to the first line is that $\braket{(\hat X^b_\alpha)^2}$ exponentially decays to its value in the vacuum state, $\braket{(\hat X^b_\alpha)^2}_{\rm ss} = 1/2$. The second and third lines describe correlations between $\hat X^b_{\Sigma(\Pi)}$ and a different quadrature, and they exponentially decay to zero. The correlation in the last line stays constant. 
Similar equations can be obtained for the elements of $\Xi^b_{YY}$. 
Thus, the steady-state solution for the covariance matrices for the Bogoliubov bosons is
\begin{align}\label{eqn: bogoliubov covariances}
& \Xi^b_{XX} = {\rm diag}\left(1,1, \vec{y}_1\cdot\vec{y}_1, \vec{y}_2\cdot\vec{y}_2, \cdots\right) \nonumber\\
& \Xi^b_{YY} = {\rm diag}\left(1,1, \vec{x}_1\cdot\vec{x}_1, \vec{x}_2\cdot\vec{x}_2, \cdots\right), \nonumber\\
& \Xi^b_{XY} = 0.
\end{align}

\subsection{Squeezing in the multilevel system}\label{subsec: multilevel squeezing}
\begin{figure}[t]\centering
\includegraphics[width=1.0\columnwidth]{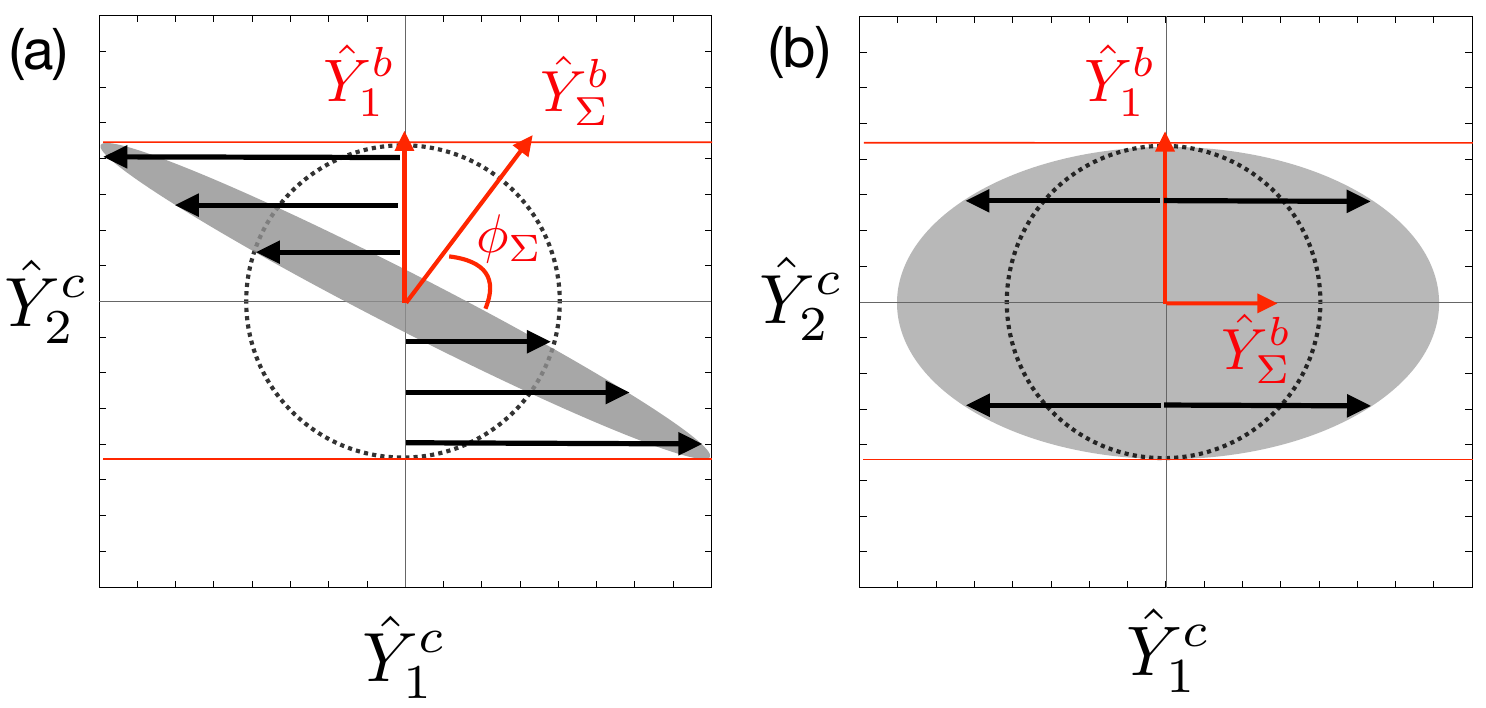}
\caption{
Visualization of the evolution of the quantum noise distribution, projected onto the $\hat{Y}^c_1$-$\hat{Y}^c_2$ plane (see text), and the emergence of squeezing and/or antisqueezing. The noise distribution in the initial coherent state is isotropic (dashed circle), and the evolution conserves the noise along $\hat{Y}^c_2$. (a) When $\phi_\Sigma\neq0$ [see Eq.~\eqref{eqn: specific choice Schwinger}], the circular noise distribution shears into an ellipse, leading to one squeezed and one antisqueezed mode. (b) When $\phi_\Sigma=0$, the circular noise distribution fattens into an ellipse if $\|\vec{y}_\Sigma\| < \|\vec{x}_\Sigma\|$, leading to one antisqueezed mode but no squeezed modes (and shrinks into a narrower ellipse with one squeezed mode if $\|\vec{y}_\Sigma\| > \|\vec{x}_\Sigma\|$). 
}
\label{fig: shearing}
\end{figure}

\begin{figure}[t]\centering
\includegraphics[width=1.0\columnwidth]{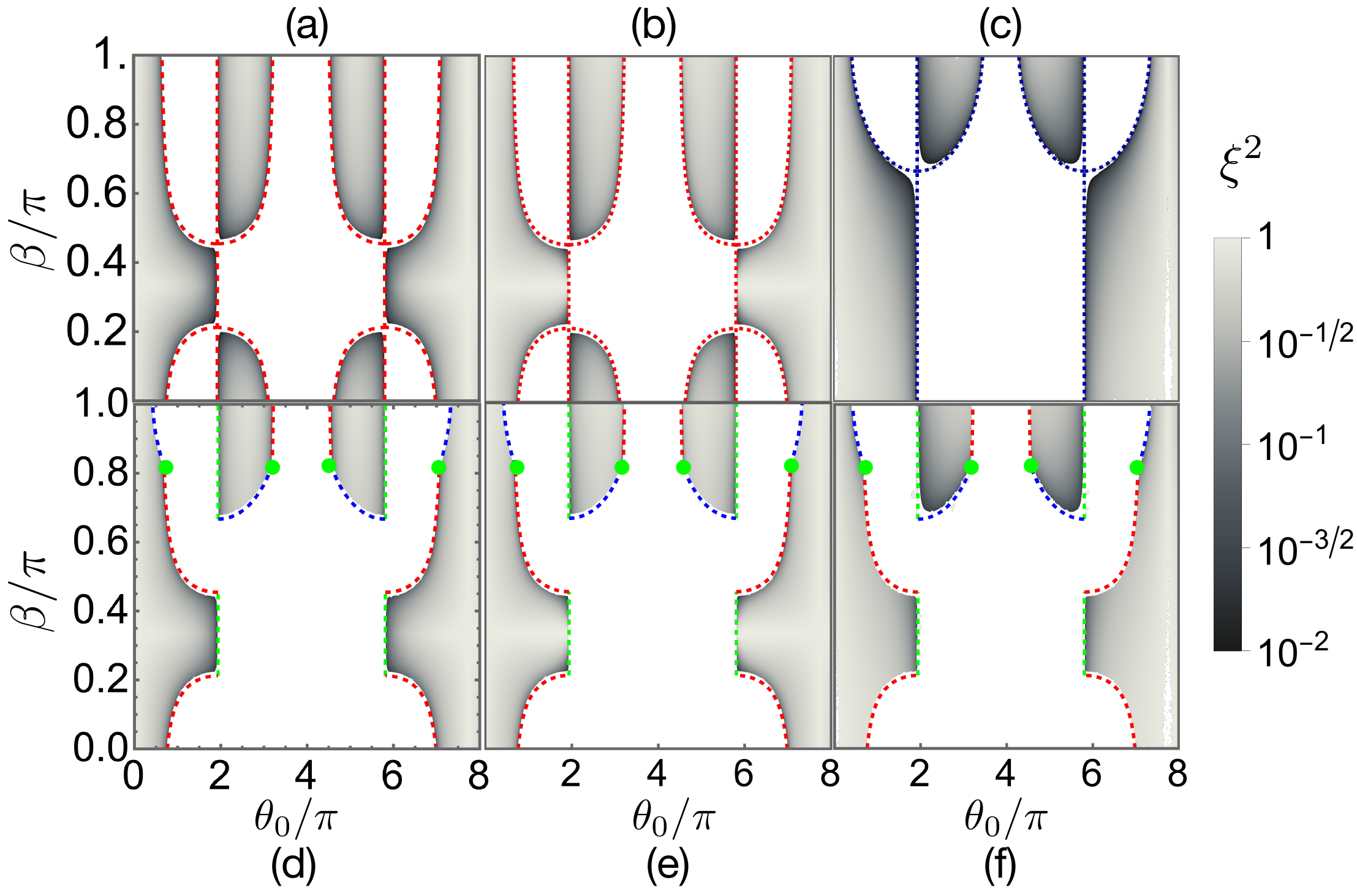}
\caption{
Steady-state squeezing for the eight-level system shown in Fig.~\ref{fig: potentials}(c). (a-b) Squeezing in a combination of the $\hat X$ and $\hat Y$ quadratures, respectively, when the system collectively emits only $\Sigma$-polarized light. (c) Squeezing when the system collectively emits only $\Pi$-polarized light. There are two squeezed modes, one in a combination of $\hat X$ quadratures and one in a combination of $\hat Y$ quadratures, and they both have the same value. Red and blue dashed lines are critical lines, and white regions are unstable.
(d-f) Squeezing when the system collectively emits light of both polarizations. The system is squeezed only in the regions where it is stable to emission of both polarizations, and the value of the squeezing in this region is the same as in (a-c).
}
\label{fig: 8 level}
\end{figure}

The steady-state covariance matrices of the Schwinger c-bosons, obtained by inverting the Bogoliubov transformation, have a nontrivial form, and host squeezed modes. Inverting the Bogoliubov transformation for the $14\times 14$ dimensional matrix, and understanding why there is squeezing, is nontrivial. However, an appropriate basis rotation of the $\ket{\mu>0}$ states makes the calculations simpler and gives a geometric understanding of the generation of squeezing (the squeezing itself is independent of the basis transformation).

\Rev{We make the basis transformation
\begin{align}
& \sum_{\mu>0} \frac{x_{\Sigma,\mu}}{\| \vec x_\Sigma \|} \hat X^c_\mu \rightarrow \hat X^c_1, \nonumber\\
& \sum_{\mu>0} \frac{x_{\Pi,\mu}}{\| \vec x_\Pi \|} \hat X^c_\mu \rightarrow \hat X^c_2, \nonumber\\
& \sum_{\mu>0} \frac{x_{1,\mu}}{\| \vec{x}_1\|} \hat X^c_\mu \rightarrow \hat X^c_3, \nonumber\\
& \sum_{\mu>0} \frac{x_{2,\mu}}{\| \vec{x}_2\|} \hat X^c_\mu \rightarrow \hat X^c_4, \nonumber\\
& \sum_{\mu>0} \frac{x_{\nu,\mu}}{\| \vec{x}_{\nu} \|} \hat X^c_\mu \rightarrow \hat X^c_{\nu+2}, \quad \nu > 2
\end{align}
where $\vec{x}_1 = \vec{y}_\Sigma - \vec{x}_\Sigma \frac{\vec{x}_\Sigma \cdot \vec{y}_\Sigma}{\| \vec{x}_\Sigma \|^2}$, $\vec{x}_2 = \vec{y}_\Pi - \vec{x}_\Pi \frac{\vec{x}_\Pi \cdot \vec{y}_\Pi}{\| \vec{x}_\Pi \|^2}$ [see Appendix~\ref{app: xy relations}], and all the $\vec{x}_{\nu}$ are orthogonal to each other.}
This basis transformation transforms the $\vec{x}_\alpha$ and $\vec{y}_\alpha$ vectors to 
\begin{align}\label{eqn: specific choice Schwinger}
& \vec{x}_\Sigma = \|\vec{x}_\Sigma\| (1, 0, 0, \cdots),\nonumber\\
& \vec{y}_\Sigma = \|\vec{y}_\Sigma\|(\cos\phi_\Sigma, \sin\phi_\Sigma, 0, \cdots),\nonumber\\
& \vec{x}_\Pi = \|\vec{x}_\Pi\| (0, 0, 1,0,0, \cdots),\nonumber\\
& \vec{y}_\Pi = \|\vec{y}_\Pi\|(0,0,\cos\phi_\Pi, \sin\phi_\Pi, 0, \cdots),\nonumber\\
& \vec{x}_1 \propto (0, 1, 0, \cdots)\nonumber\\
& \vec{y}_1 \propto (-\sin\phi_\Sigma,\cos\phi_\Sigma, 0, \cdots), \nonumber\\
& \vec{x}_2 \propto (0,0,0, 1, 0, \cdots),\nonumber\\
& \vec{y}_2 \propto (0,0,-\sin\phi_\Pi, \cos\phi_\Pi, 0, \cdots),\nonumber\\
& x_{\mu,\nu} = y_{\mu,\nu} = \delta_{\mu,\nu+2}\quad (\mu > 2).
\end{align}
\Rev{After this transformation, the relation between the Bogoliubov b-bosons and the Schwinger c-bosons is simpler. In particular, for the $\hat{X}$ variables we find
\begin{align}
& \hat X^b_\Sigma = \frac{\| x_\Sigma\|}{\sqrt{\vec{x}_\Sigma \cdot \vec{y}_\Sigma}}\hat X^c_1 ,\nonumber\\
& \hat X^b_\Pi = \frac{\| x_\Pi\|}{\sqrt{\vec{x}_\Pi \cdot \vec{y}_\Pi}}\hat X^c_2 , \nonumber\\
& \hat X^b_1 = \|\vec{y}_1\| (\sin\phi_\Sigma \hat X^c_1 - \cos\phi_\Sigma \hat X^c_3) 
\nonumber\\
& \hat X^b_2 = \|\vec{y}_2\| (\sin\phi_\Pi \hat X^c_2 - \cos\phi_\Pi \hat X^c_4) 
\nonumber\\
& \hat X^b_{\mu> 2} = \hat X^c_{\mu+2},
\end{align}
where $\cos\phi_\alpha = \frac{\vec{x}_\alpha \cdot \vec{y}_\alpha}{\|\vec{x}_\alpha\| \|\vec{y}_\alpha\|}$, $\vec{y}_1 = \vec{x}_\Sigma - \vec{y}_\Sigma \frac{\vec{x}_\Sigma \cdot \vec{y}_\Sigma}{\| \vec{y}_\Sigma \|^2}$, and $\vec{y}_2 = \vec{x}_\Pi - \vec{y}_\Pi \frac{\vec{x}_\Pi \cdot \vec{y}_\Pi}{\| \vec{y}_\Pi \|^2}$ [see Appendix~\ref{app: xy relations}].} 
This basis transformation is useful because it block diagonalizes the covariance matrix of the Schwinger $c$-bosons at all times. This is because in this basis $\hat{b}_\Sigma$ and $\hat{b}_1$ only depend on $\hat{c}_{1,2}$, $\hat{b}_\Pi$ and $\hat{b}_2$ only depend on $\hat{c}_{3,4}$, and $\hat{b}_{\mu\geq3}=\hat{c}_{\mu+2}$.
This facilitates the visualization of each pair of squeezed and corresponding antisqueezed modes in a two-dimensional space that is independent of the other squeezed and antisqueezed modes, as we explain below.

Equation~\eqref{eqn: bogoliubov dynamics} shows that the $\hat X$ quadratures evolve independently from the $\hat Y$ quadratures, so we will consider them separately, focusing first on the $\hat Y$ quadratures and applying a similar argument to the $\hat X$ quadratures. Because of the structure of the basis choice in Eq.~\eqref{eqn: specific choice Schwinger}, the covariances of $\hat Y^c_1$ and $\hat Y^c_2$ have coupled master equations, the covariances of $\hat Y^c_3$ and $\hat Y^c_4$ have coupled master equations, and all other covariances evolve independently. Thus, we will focus on the evolution of $\hat Y^c_1$ and $\hat Y^c_2$ first.

The two most important elements to understand the evolution of these covariances are the following.
(I) The noise along $\hat Y^b_\Sigma \propto (\cos\phi_\Sigma \hat Y^c_1 + \sin\phi_\Sigma \hat Y^c_2)$ evolves towards its vacuum value, as discussed previously;
(II) the noise in $\hat Y^c_2$ is conserved, since $\hat Y^c_2$ commutes with $\hat{\mathscr{D}}^-_\Sigma$ and $\hat{\mathscr{D}}^-_\Pi$ [see Eqs.~\eqref{eq:8l_BogPiSigma_def},~\eqref{eq:8l_Bogconserved_def}, and~\eqref{eqn: specific choice Schwinger}].
Note that the initial noise of $\hat Y^c_1$ and $\hat Y^c_2$ in the initial coherent state is equal, i.e., the noise has a circular distribution. In the general case where $\phi_\Sigma \neq 0$, the conservation of $\hat Y^c_2$ sets a constraint on $\hat Y^b_\Sigma$ resulting in an evolution that shears the circle into an ellipse as shown in Fig.~\ref{fig: shearing}(a), which leads to one squeezed and one antisqueezed mode. Similar shearing on the $\hat Y^c_3$-$\hat Y^c_4$ plane due to emission of $\Pi$ polarization leads again to one squeezed and one antisqueezed mode, and a similar process happens in the $\hat X^c$ quadratures. In total, there are four squeezed and four antisqueezed modes. \Rev{Their values can be obtained from diagonalizing the Schwinger bosons' covariance matrix. In the transformed basis, for example,
\begin{widetext}
\begin{equation}\label{eqn: Xic multilevel}
\Xi^c_{XX} = \left( \begin{array}{cccccc}
\frac{\vec{x}_\Sigma \cdot \vec{y}_\Sigma}{\| x_\Sigma\|^2} & 0 & \frac{\sqrt{\|x_\Sigma\|^2\|y_\Sigma\|^2 - (\vec{x}_\Sigma\cdot\vec{y}_\Sigma)^2}}{\|x_\Sigma\|^2} & \cdots &&\\
0 & \frac{\vec{x}_\Pi \cdot \vec{y}_\Pi}{\| x_\Pi\|^2} & 0 & \cdots &&\\
\frac{\sqrt{\|x_\Sigma\|^2\|y_\Sigma\|^2 - (\vec{x}_\Sigma\cdot\vec{y}_\Sigma)^2}}{\|x_\Sigma\|^2} & 0 & 
\frac{\|y_\Sigma\|^2}{(\vec{x}_\Sigma\cdot\vec{y}_\Sigma)} + \frac{\|x_\Sigma\|^2\|y_\Sigma\|^2}{(\vec{x}_\Sigma\cdot\vec{y}_\Sigma)^2} - \frac{\vec{x}_\Sigma\cdot\vec{y}_\Sigma}{\|x_\Sigma\|^2}
 & \cdots &&\\
0 & 0 & \cdots &&&\\
0 & 0 & 0 & 0 & 1 & \cdots\\
\vdots &&&&& \ddots
\end{array}\right)
\end{equation}
\end{widetext}
The covariance matrix elements of $\Xi^c_{YY}$ can be similarly computed. The nontrivial eigenvalues of $\Xi^c_{XX}$ and $\Xi^c_{YY}$ are
\begin{widetext}
\begin{align}\label{eqn: squeezing}
\xi^2_{X,\alpha,\rm sq} =& \frac{ \vec{y}_\alpha\cdot\vec{y}_\alpha }{\vec{x}_\alpha\cdot\vec{y}_\alpha } \left(1+\frac{\vec{x}_\alpha\cdot\vec{x}_\alpha}{\vec{x}_\alpha\cdot\vec{y}_\alpha}\right) - \sqrt{
\left(\frac{\vec{y}_\alpha\cdot\vec{y}_\alpha}{\vec{x}_\alpha\cdot\vec{y}_\alpha}\right)^2\left(1+\frac{\vec{x}_\alpha\cdot\vec{x}_\alpha}{\vec{x}_\alpha\cdot\vec{y}_\alpha}\right)^2 - 4\frac{\vec{y}_\alpha\cdot\vec{y}_\alpha}{\vec{x}_\alpha\cdot\vec{y}_\alpha}
} , \nonumber\\
\xi^2_{Y,\alpha,\rm sq} =& \frac{ \vec{x}_\alpha\cdot\vec{x}_\alpha }{\vec{x}_\alpha\cdot\vec{y}_\alpha } \left(1+\frac{\vec{y}_\alpha\cdot\vec{y}_\alpha}{\vec{x}_\alpha\cdot\vec{y}_\alpha}\right) - \sqrt{
\left(\frac{\vec{x}_\alpha\cdot\vec{x}_\alpha}{\vec{x}_\alpha\cdot\vec{y}_\alpha}\right)^2\left(1+\frac{\vec{y}_\alpha\cdot\vec{y}_\alpha}{\vec{x}_\alpha\cdot\vec{y}_\alpha}\right)^2 - 4\frac{\vec{x}_\alpha\cdot\vec{x}_\alpha}{\vec{x}_\alpha\cdot\vec{y}_\alpha}
} , \nonumber\\
\xi^2_{X,\alpha,\rm anti-sq} =& \frac{ \vec{y}_\alpha\cdot\vec{y}_\alpha }{\vec{x}_\alpha\cdot\vec{y}_\alpha } \left(1+\frac{\vec{x}_\alpha\cdot\vec{x}_\alpha}{\vec{x}_\alpha\cdot\vec{y}_\alpha}\right) + \sqrt{
\left(\frac{\vec{y}_\alpha\cdot\vec{y}_\alpha}{\vec{x}_\alpha\cdot\vec{y}_\alpha}\right)^2\left(1+\frac{\vec{x}_\alpha\cdot\vec{x}_\alpha}{\vec{x}_\alpha\cdot\vec{y}_\alpha}\right)^2 - 4\frac{\vec{y}_\alpha\cdot\vec{y}_\alpha}{\vec{x}_\alpha\cdot\vec{y}_\alpha}
}, \nonumber\\
\xi^2_{Y,\alpha,\rm anti-sq} =& \frac{ \vec{x}_\alpha\cdot\vec{x}_\alpha }{\vec{x}_\alpha\cdot\vec{y}_\alpha } \left(1+\frac{\vec{y}_\alpha\cdot\vec{y}_\alpha}{\vec{x}_\alpha\cdot\vec{y}_\alpha}\right) + \sqrt{
\left(\frac{\vec{x}_\alpha\cdot\vec{x}_\alpha}{\vec{x}_\alpha\cdot\vec{y}_\alpha}\right)^2\left(1+\frac{\vec{y}_\alpha\cdot\vec{y}_\alpha}{\vec{x}_\alpha\cdot\vec{y}_\alpha}\right)^2 - 4\frac{\vec{x}_\alpha\cdot\vec{x}_\alpha}{\vec{x}_\alpha\cdot\vec{y}_\alpha}
}.
\end{align}
\end{widetext}
$\xi^2_{X,\alpha,\rm sq}$ and $\xi^2_{Y,\alpha,\rm sq}$ are the values of squeezing, and $\xi^2_{X,\alpha,\rm anti-sq}$ and $\xi^2_{Y,\alpha,\rm anti-sq}$ are the antisqueezing values.
}

The special case $\phi_\Sigma=0$ is qualitatively different from the general case of $\phi_\Sigma \neq 0$ [see Fig.~\ref{fig: shearing}(b)]. In this case, we have $\hat Y^b_\Sigma = \sqrt{\frac{||\vec{y}_{\Sigma}||}{\|\vec{x}_\Sigma\|}} \hat Y^c_1$ and thus its evolution towards the vacuum value is unconstrained by $\hat{Y}^c_2$. Therefore, the noise in $\hat Y^b_\Sigma$ increases to its vacuum state value if $\|\vec{y}_\Sigma\| < \|\vec{x}_\Sigma\|$ leading to antisqueezing in $\hat Y^c_1$, and the noise in $\hat Y^b_\Sigma$ decreases to its vacuum state value if $\|\vec{y}_\Sigma\| > \|\vec{x}_\Sigma\|$ leading to squeezing in $\hat Y^c_1$. The opposite happens in $\hat X^b_\Sigma$ and $\hat X^c_1$. Because of this, the number of squeezed modes is reduced by 1 compared to the general case $\phi_\Sigma \neq 0$. This is essentially what happens in the two-level system of Sec.~\ref{sec: two-level}, where we had that $\vec{x}_R$ was parallel to $\vec{y}_R$ (instead of $\vec{x}_\Sigma$ and $\vec{y}_\Sigma$).

Figure~\ref{fig: 8 level} plots the steady-state values of the squeezing vs $\theta_0$ and $\beta$.
Figures~\ref{fig: 8 level}(a-b) show the value of the squeezing in two modes if the system collectively emitted $\Sigma$-polarized light only. Figure~\ref{fig: 8 level}(c) shows the value of the squeezing if the system collectively emitted $\Pi$-polarized light only. In the latter case, two modes are squeezed but the amount of squeezing in both modes is the same, so we plot them together. The white regions are unstable to quantum fluctuations, and critical lines (dashed) separate the stable and unstable regions (the small white gaps between the critical lines and the gray regions are due to truncating the squeezing at $10^{-2}$). Note that emission of $\Sigma$- and $\Pi$-polarized light have different regions of stability. For example, the system may be stable to emission of $\Sigma$-polarized light, but unstable to emission of $\Pi$-polarized light, or vice versa. Figures~\ref{fig: 8 level}(d-f) plot the squeezing in the same four modes as Figs.~\ref{fig: 8 level}(a-c), but only in the region where the system is stable to emission of both polarizations, which is the physically relevant case. The squeezing due to emission of $\Sigma$-polarized light approaches zero near the red lines, the squeezing due to emission of $\Pi$-polarized light approaches zero near the blue lines, and squeezing in all four modes approaches zero near the green lines and points.

\subsection{Finite-size corrections in the steady state}

As in the two-level system [Sec.~\ref{subsec: two-level correlations}], the best squeezing reachable near the critical point is limited by $N$. Here, we calculate the finite-size corrections to the steady-state squeezing by including the higher-order terms in the HP approximation.

Near any critical point in a multilevel system, $\vec{x}_\alpha \cdot \vec{y}_\alpha$ approaches zero, which means generally that the angle $\phi_\alpha$ [Eq.~\eqref{eqn: specific choice Schwinger}] between them approaches $\pi/2$ (for the two-level-like case [see Fig.~\ref{fig: shearing}(b)], $\|\vec{y}_\alpha\|$ approaches zero near the critical point, and the arguments below do not apply). From Fig.~\ref{fig: shearing}(a), we see that for $\phi_\alpha \approx \pi/2$, the squeezed variable is approximately $\hat Y^c_{\alpha,\rm sq} \equiv \frac{\sum_\mu y_{\alpha,\mu}^{\phantom\dagger} \hat Y^c_\mu}{\| \vec{y}_\alpha\|} \propto \hat Y^b_\alpha$. Similarly, $\hat X^c_{\alpha,\rm sq} \equiv \frac{\sum_\mu x_{\alpha,\mu}^{\phantom\dagger} \hat X^c_\mu}{\| \vec{x}_\alpha\|} \propto \hat X^b_\alpha$, and the antisqueezed quadratures are $\hat{ X}^c_{\alpha,\rm antisq} = \frac{\sum_\mu y_{\alpha,\mu}^{\phantom\dagger} \hat X^c_\mu}{\| \vec{y}_\alpha\|}$ and $\hat{Y}^c_{\alpha,\rm antisq} = \frac{\sum_\mu x_{\alpha,\mu}^{\phantom\dagger} \hat Y^c_\mu}{\| \vec{x}_\alpha\|}$, $\alpha = \Sigma,\Pi$. Since $\vec{x}_\alpha \cdot \vec{y}_\alpha$ approaches zero, we expand the squeezing and antisqueezing in powers of $\vec{x}_\alpha \cdot \vec{y}_\alpha$. Focusing on only one pair of these variables, $\hat X^c_{\alpha,\rm sq}$ and $\hat{X}^c_{\alpha,\rm anti-sq}$, as an example, their steady-state values are [see Eq.~\eqref{eqn: squeezing}]
\begin{align}\label{eqn: multilevel squeezing near critical pt}
& \xi^2_{X,\alpha, \rm sq} \approx 2\frac{\vec{x}_\alpha \cdot \vec{y}_\alpha}{\vec{x}_\alpha \cdot \vec{x}_\alpha} + O(1/N)\nonumber\\
& \xi^2_{X,\alpha,\rm anti-sq} \approx \xi^2_{Y,\alpha,\rm anti-sq} \approx 2\frac{(\vec{x}_\alpha \cdot \vec{x}_\alpha)(\vec{y}_\alpha \cdot \vec{y}_\alpha)}{(\vec{x}_\alpha \cdot \vec{y}_\alpha)^2} + O(1/N).
\end{align}

The $O(1/N)$ terms in both equations arise due to the higher-order terms, i.e.,~the $g_{\alpha,\mu\nu>0}$ terms, in the master equation. As the critical point is approached, those $O(1/N)$ terms increase proportionally to $(\xi^2_{X,\alpha,\rm anti-sq})/(\vec{x}_\alpha\cdot\vec{y}_\alpha)$. This in particular affects the squeezing:
\begin{equation}\label{eqn: multilevel squeezing near critical pt}
\xi^2_{X,\alpha, \rm sq} \approx \frac{\vec{x}_\alpha \cdot \vec{y}_\alpha}{\vec{x}_\alpha \cdot \vec{x}_\alpha} + \underbrace{\frac{A}{N(\vec{x}_\alpha\cdot\vec{y}_\alpha)^3}}_{\rm finite\ size} ,
\end{equation}
where $A$ is some constant. Thus, the squeezing does not decrease monotonically; instead it increases as the critical point is approached. The optimal value of $\vec{x}_\alpha\cdot\vec{y}_\alpha$ scales as $\vec{x}_\alpha\cdot\vec{y}_\alpha \propto N^{-1/4}$, and the optimum squeezing also scales $\propto N^{-1/4}$ [see also Ref.~\cite{sundar2023squeezing}].

\section{Discussion}\label{sec: discussion}
We described a method to produce a collective four-mode squeezed state of matter using the interplay of driving and dissipation in a cavity. For the model considered, there are two main differences in the nature of the squeezing dynamics in the multilevel systems as compared to the well-known case of two-level systems.

First, driven-dissipative dynamics in two-level systems generate only one squeezed mode, whereas dynamics in multilevel systems can generically produce up to two squeezed modes per polarization. In Ref.~\cite{sundar2023squeezing}, we studied cases when only one cavity polarization is relevant and explained that squeezing emerges from shearing perpendicular to two conserved spin variables. Here we generalized the analysis for the more general case when two polarizations are in play.

The second difference between two-level and multilevel systems is the finite-size scaling of the best squeezing near the critical points. Near the critical point, the squeezing gets an admixture of the antisqueezing, which limits the best squeezing achievable. The antisqueezing increases faster in the multilevel system than in the two-level system, as the critical point is approached.
Therefore, the scaling of the best squeezing in a multilevel system is usually worse ($\propto N^{-1/4}$) than a two-level system ($\propto N^{-1/3}$).

We have focused on a specific level structure and type of initial conditions. However, there is still a large parameter space to explore the dynamics and squeezing generation of multilevel atoms. 
While our results hold for cases with a single ground and excited manifold, one might consider more general level structures with multiple hyperfine ground or excited manifolds, which could be relevant to alkali-metal atoms. These more general cases might show richer behaviors. We note that our formalism can be straightforwardly applied to these richer cases as well.

For the sake of simplicity, we only considered 
cases when the mean-field dynamics starts at a stable stationary state. However, extending the analysis to more general situations where the mean-field dynamics is nontrivial could lead to more interesting steady states and phases. 
For example, quantum fluctuations may drive an initially unstable state towards a mixture of stable macroscopic steady states which could be entangled. Furthermore, the large number of steady states and unstable regions anticipates a rich phase diagram with superradiant to normal transitions analogous to two-level atoms, as well as potentially other types of transitions such as superradiant to superradiant.

While we considered the generation of squeezing in a system with only coherent driving and collective emission of light, the cavity can also mediate elastic interactions between the atoms via exchange of photons~\cite{norcia2018cavity,Muniz2020,barberena2019driven}. The interplay between elastic interactions and the dissipation could be an interesting question for the evolution and finite-size scaling of the squeezing. The effects of other decoherence sources such as spontaneous emission or dephasing on the squeezing, as well as the effect of experimental details such as inhomogeneous couplings, are also important questions to address in future work.

Finally, in Ref.~\cite{sundar2023squeezing} we showed that it is possible to prepare a squeezed state and rotate it into a state that is dark to emission on one polarization by taking advantage of the conserved quadratures.
That analysis can be extended to the case of two polarizations, such that the four squeezed modes discussed here can be preserved in dark states.
Furthermore, since atoms with many levels will contain many conserved quadratures ($\ell-3$), it is in principle possible to create squeezing, store it in a conserved quadrature, then create squeezing again and store it in the remaining conserved quadratures. This would allow the creation of multilevel spin states with many squeezed directions which might be useful for multi-parameter quantum sensing protocols~\cite{kaubruegger2023optimal}. 

\begin{acknowledgments}
We thank Jeremy T. Young, Dylan Young, and James K. Thompson for helpful discussions and feedback. This work is supported by the VBFF,  AFOSR grants FA9550-18-1-0319,  by the NSF JILA-PFC PHY-2317149, QLCI-OMA-2016244, by the U.S. Department of Energy, Office of Science, National Quantum Information Science Research Centers Quantum Systems Accelerator, and by NIST.
\end{acknowledgments}

\bibliography{bibliography}

\appendix
\begin{widetext}

\section{Deriving the effective multilevel spin model}\label{app: effective master equation}
The dynamics of the atom-light system is modeled by the Lindblad master equation:
\begin{equation}\label{eqn: master}
\hbar\frac{d\rho}{dt} = -i[\hat H_{\rm tot}, \rho] + \mathcal{L}_{\rm cav}[\rho].
\end{equation}
Here, $\hat H_{\rm tot} = \hat H_A + \hat H_L + \hat H_{AL}$ is a Hamiltonian including contributions from the atoms, cavity modes, atom-light coupling, and external driving:
\begin{align}\label{eqn: H spin-photon}
& \hat H_A = \hbar\omega \hat n_e,\nonumber\\
& \hat H_L = \sum_{\alpha} \hbar\omega \ha_\alpha\+ \ha_\alpha^{\phantom\dagger} + \frac{i\hbar\epsilon_\alpha}{2}(\ha_\alpha\+ e^{i\omega t} - \ha_\alpha^{\phantom\dagger} e^{-i\omega t}),\nonumber\\
& \hat H_{AL} = \hbar g \sum_\alpha \ha_\alpha \hat{D}_\alpha^+ + {\rm H.c.} ,
\end{align}
and $\mathcal{L}_{\rm cav}[\rho]$ describes the dissipation terms due to leakage of photons out of the cavity at rate $\kappa$:
\begin{equation}\label{eqn: Lcav}
\mathcal{L}_{\rm cav}[\rho] = \hbar\kappa\sum_{\alpha} \left(\ha_\alpha^{\phantom\dagger} \rho \ha_\alpha\+ - \frac{1}{2}\ha_\alpha\+\ha_\alpha^{\phantom\dagger} \rho - \frac{1}{2}\rho\ha_\alpha\+\ha_\alpha^{\phantom\dagger}\right) .
\end{equation}
In the above equations, $\hat n_e$ is the occupation in the excited manifold, and $\ha_\alpha$ annihilates a photon in the cavity mode with polarization $\alpha$.

It is convenient to move to a rotating frame that rotates at the atomic and photon frequency $\omega$. In this frame, the atomic angular frequency, cavity frequency, and laser frequency are shifted by $\omega$, yielding the Hamiltonian
\begin{equation}
\hat H_{\rm tot} = \sum_\alpha \frac{i\hbar\epsilon_\alpha}{2}(\ha_\alpha\+ - \ha_\alpha^{\phantom\dagger}) + \hbar g \left( \ha_\alpha \hat{D}_\alpha^+ + {\rm H.c.} \right),
\end{equation}
and the same Lindblad jump operators as before.

The master equation for the photon operators is
\begin{equation}
\partial_t \braket{\ha_\alpha} = \braket{ -\frac{\kappa}{2} \ha_\alpha - i g \hat D_\alpha^- + \frac{\epsilon_\alpha}{2} }.
\end{equation}
Assuming the bad cavity limit, $\kappa \gg g\sqrt{N}$, the photons' evolution follows the spins:
\begin{equation}\label{eqn: a in terms of spins}
\ha_\alpha \rightarrow \frac{i\epsilon_\alpha + 2g \hat D_\alpha^-}{i\kappa}.
\end{equation}
We adiabatically eliminate the photons by substituting Eq.~\eqref{eqn: a in terms of spins} into Eqs.~\eqref{eqn: H spin-photon} and~\eqref{eqn: Lcav}, and obtain
\begin{equation}\label{eqn: Heff and Leff}
\hat H_{\rm eff} = 0,\quad \hat{L}_{\rm eff} = \frac{\sqrt{\hbar\kappa}(i\epsilon_\alpha + 2g \hat D_\alpha^-)}{i\kappa}.
\end{equation}
The Lindbladian equation due to $\hat{L}_{\rm eff}$ is identical to the master equation due to $\hat H_{\rm drive}$ [Eq.~(\ref{eqn: H})] and $\mathcal{L}$ [Eq.~\eqref{eqn: L}], with $\Omega_\alpha = \frac{2\epsilon_\alpha g}{\kappa}$ and $\Gamma = \frac{4g^2}{\kappa}$.

\section{Higher-order effects on the stability of the mean-field state}\label{app subsec: higher-order effect on stability}
In the main text, we derived the leading-order condition for the stability of the mean-field state by looking at the evolution of two-body observables and arguing that $\braket{\hat{\mathscr{D}}^+_{\alpha_i}\hat{\mathscr{D}}^-_{\alpha_j}}=0$ to leading order at the steady state. We assumed that $\lambda_{ij}$ is constant, $\braket{\hat{\mathscr{D}}^-_\alpha} = 0$, and that the cumulant approximation is valid. 
Here we use the HP approximation to show that Eq.~(\ref{eqn: eqns in cumulant approx}) is indeed correct to the order in $N$ considered, and argue that higher-order corrections become relevant only at time scales of order $O(1/\sqrt{N}\Gamma)$.

In the HP approximation, $\lambda_{ij} \equiv \braket{[\hat{\mathscr{D}}^-_{\alpha_i}, \hat{\mathscr{D}}^+_{\alpha_j}]} = 2N\sqrt{(\vec{x}_{\alpha_i} \cdot \vec{y}_{\alpha_i})(\vec{x}_{\alpha_j} \cdot \vec{y}_{\alpha_j})} \braket{[\hat b_{\alpha_i}^{\phantom\dagger}, \hat b_{\alpha_j}\+]} + O(\sqrt{N})$. The leading-order term in $\lambda_{ij}$ is proportional to the commutator of $[\hat b_{\alpha_i}^{\phantom\dagger}, \hat b_{\alpha_j}\+]$, which is a constant. If the polarization basis is chosen such that the Bogoliubov b-bosons commute, then $\lambda_{ij} = 2N(\vec{x}_{\alpha_i} \cdot \vec{y}_{\alpha_i})\delta_{ij} + O(\sqrt{N})$.

Similarly, in the HP expansion we have the correlation $\braket{\hat{\mathscr{D}}^+_{\alpha_i}\hat{\mathscr{D}}^-_{\alpha_j}} = 2N\sqrt{(\vec{x}_{\alpha_i} \cdot \vec{y}_{\alpha_i})(\vec{x}_{\alpha_j} \cdot \vec{y}_{\alpha_j})} \braket{\hat b_{\alpha_i}\+ \hat b_{\alpha_j}^{\phantom\dagger}} + O(\sqrt{N})$. The master equation for $\braket{\hat b_{\alpha_i}\+ \hat b_{\alpha_j}^{\phantom\dagger}}$ is
\begin{equation}
\partial_t \braket{\hat b_{\alpha_i}\+\hat b_{\alpha_j}^{\phantom\dagger}} = \sum_\alpha N\Gamma\, \vec{x}_\alpha\cdot\vec{y}_\alpha \braket{ [\hat b_\alpha\+, \hat b_{\alpha_i}\+\hat b_{\alpha_j}^{\phantom\dagger}] \hat b_\alpha + \hat b_\alpha\+ [\hat b_{\alpha_i}\+\hat b_{\alpha_j}^{\phantom\dagger}, \hat b_\alpha] } + O(\sqrt{N}\Gamma).
\end{equation}
Expanding the commutators and inserting the definition of $\lambda_{ij}$ yields
\begin{equation}
\partial_t \braket{\hat b_{\alpha_i}\+\hat b_{\alpha_j}^{\phantom\dagger}} = -\frac{\Gamma}{2} \sum_\alpha \sqrt{\vec{x}_\alpha\cdot\vec{y}_\alpha} \left( \frac{\lambda_{\alpha_j,\alpha}}{\sqrt{ \vec{x}_{\alpha_j}\cdot\vec{y}_{\alpha_j}}} \braket{\hat b_{\alpha_i}\+ \hat b_\alpha^{\phantom\dagger}} + \frac{\lambda_{\alpha,\alpha_i}}{\sqrt{ \vec{x}_{\alpha_i}\cdot\vec{y}_{\alpha_i}}} \braket{\hat b_\alpha\+ \hat b_{\alpha_j}^{\phantom\dagger}} \right) + O(\sqrt{N}\Gamma).
\end{equation}
Reverting from the Bogoliubov b-boson correlator $\hat b\+_\alpha \hat b_\beta^{\phantom\dagger}$ back to the spin correlator $\frac{1}{2N\sqrt{(\vec{x}_{\alpha} \cdot \vec{y}_{\alpha})(\vec{x}_{\beta} \cdot \vec{y}_{\beta})}} \braket{\hat{\mathscr{D}}^+_{\alpha}\hat{\mathscr{D}}^-_{\beta}}$ gives
\begin{equation}
\partial_t \braket{\hat{\mathscr{D}}^+_{\alpha_i}\hat{\mathscr{D}}^-_{\alpha_j}} = -\frac{\Gamma}{2} \sum_\alpha \left( \lambda_{\alpha_j,\alpha} \braket{\hat{\mathscr{D}}^+_{\alpha_i} \hat{\mathscr{D}}^-_\alpha} + \lambda_{\alpha,\alpha_i} \braket{\hat{\mathscr{D}}^+_\alpha \hat{\mathscr{D}}^-_{\alpha_j}} \right) + O(\sqrt{N}\Gamma).
\end{equation}
The leading terms give Eq.~\eqref{eqn: eqns in cumulant approx}, and drive dynamics on the time scale of $1/(N\Gamma)$, since $\lambda_{ij} \sim O(N)$. The sub-leading terms can drive dynamics on the time scale of $O(1/\sqrt{N}\Gamma)$, which is much longer than the $O(1/N\Gamma)$ timescale of the dominant dynamics.

\section{Relation between the $\vec{x}$ and $\vec{y}$ vectors}\label{app: xy relations}
Here, we show that for the state $\ket{\Psi} \equiv \ket{\Psi(\alpha_0,\theta_0;
\vec{\beta})}$ considered here, we can make an appropriate basis choice such that all the $\vec{x}$ vectors are orthogonal to each other, and all the $\vec{y}$ vectors are orthogonal to each other, and also that $\vec{x}_\Sigma \cdot \vec{y}_\Pi = \vec{x}_\Pi \cdot \vec{y}_\Sigma = 0$. We will do this in three steps.

\subsection{$\vec{x}_\Sigma$ and $\vec{y}_\Sigma$ are orthogonal to $\vec{x}_\Pi$ and $\vec{y}_\Pi$}
The jump operators are $\hat{\mathscr{D}}^-_\alpha = \hat{\mathscr{D}}^x_\alpha - i \hat{\mathscr{D}}^y_\alpha$, where their real and imaginary parts are $\hat{\mathscr{D}}^x_\alpha = \sqrt{N} \sum_{\mu>0} x_{\alpha,\mu} \hat X^c_\mu$ and $\hat{\mathscr{D}}^y_\alpha = \sqrt{N} \sum_{\mu>0} y_{\alpha,\mu} \hat Y^c_\mu$.

\noindent \textbf{Statement 1:} $\vec{x}_\Sigma\cdot\vec{y}_\Pi = 0$ and $\vec{x}_\Pi\cdot\vec{y}_\Sigma = 0$.\\
\textbf{Proof:} 
Due to the choice of the quantization axis, we have that $\braket{ \Psi \vert [\hat{\mathscr{D}}^-_\Sigma, \hat{\mathscr{D}}^+_\Pi] \vert \Psi} = 0$. We also have the operator identity $[\hat{\mathscr{D}}^-_\Sigma, \hat{\mathscr{D}}^-_\Pi] = 0$. Applying the HP approximation and the standard commutation relations for $\hat X^c_\mu$ and $\hat Y^c_\mu$ leads to Statement 1.

\noindent \textbf{Statement 2:} $\vec{x}_\Sigma\cdot\vec{x}_\Pi = 0$.\\
\textbf{Proof:} 
Since $\ket{\Psi}$ is the vacuum of $\hat c_{\mu \neq 0}$, therefore $\braket{\Psi \vert \hat X^c_\mu \hat X^c_\nu \vert \Psi} = \frac{1}{2}\delta_{\mu\nu}$, and
\begin{equation}\label{eqn: dSigma dPi}
\braket{\Psi \vert \hat{\mathscr{D}}^x_\Sigma \hat{\mathscr{D}}^x_\Pi \vert \Psi} = \frac{N}{2}\vec{x}_\Sigma \cdot \vec{x}_\Pi.
\end{equation}
Since $\ket{\Psi}$ is a coherent state, Eq.~\eqref{eqn: dSigma dPi} can be explicitly calculated, and it evaluates to zero. This proves Statement 2.

\noindent \textbf{Statement 3:} $\vec{y}_\Sigma\cdot\vec{y}_\Pi = 0$.\\
\textbf{Proof:} 
We note that
\begin{align}
& \braket{ \Psi \vert \hat{\mathscr{D}}^x_\Sigma \hat{\mathscr{D}}^x_\Pi \vert \Psi} = \sum_i \braket{ \Psi \vert \hat{d}^x_{i,\Sigma} \hat{d}^x_{i,\Pi} \vert \Psi},\nonumber\\
& \braket{ \Psi \vert \hat{\mathscr{D}}^y_\Sigma \hat{\mathscr{D}}^y_\Pi \vert \Psi} = \sum_i \braket{ \Psi \vert \left(\hat{d}^y_{i,\Sigma}-\frac{\Omega_\Sigma}{N\Gamma}\right) \left( \hat{d}^y_{i,\Pi}-\frac{\Omega_\Pi}{N\Gamma}\right) \vert \Psi}.
\end{align}
The terms with $i\neq j$ are zero, due to Eq.~\eqref{eqn: mf steady state}. Next, using the operator identity $\hat{d}^y_{i,\Sigma}\hat{d}^y_{i,\Pi} = \hat{d}^x_{i,\Sigma}\hat{d}^x_{i,\Pi}$ the fact that we do not need a $\Pi$-polarized drive, $\Omega_\Pi=0$, and the value of $\Omega_\Sigma$ from Eq.~\eqref{eqn: Omega steady state}, we find that $\braket{ \Psi \vert \hat{\mathscr{D}}^y_\Sigma \hat{\mathscr{D}}^y_\Pi \vert \Psi} = \braket{ \Psi \vert \hat{\mathscr{D}}^x_\Sigma \hat{\mathscr{D}}^x_\Pi \vert \Psi} = 0$. Applying the HP approximation, this proves Statement 3.

\noindent \textbf{Statement 4:} Both $\vec{x}_\Sigma$ and $\vec{y}_\Sigma$ are orthogonal to both $\vec{x}_\Pi$ and $\vec y_\Pi$.\\
\textbf{Proof}: This follows from Statements 1, 2, and 3.

\subsection{$\vec{x}_{\alpha=\Sigma,\Pi}$ are orthogonal to $\vec{x}_{\mu \neq \Sigma,\Pi}$, and analogously for the $\vec{y}$ vectors.}
The Bogoliubov operators $\hat b_{\mu\neq\Sigma,\Pi}$ are independent of $\hat b_\Sigma$ and $\hat b_\Pi$, i.e., $[\hat b_{\mu\neq\Sigma,\Pi}, \hat b_{\alpha=\Sigma,\Pi}] = [\hat b_{\mu\neq\Sigma,\Pi}, \hat b_{\alpha=\Sigma,\Pi}\+] = 0$. This leads to the fact that $\vec{x}_{\alpha=\Sigma,\Pi}$ are orthogonal to $\vec{x}_{\mu \neq \Sigma,\Pi}$, and also that $\vec{y}_{\alpha=\Sigma,\Pi}$ are orthogonal to $\vec{y}_{\mu \neq \Sigma,\Pi}$.

A simple choice is to make $\vec{x}_1$ and $\vec{y}_1$ be coplanar with $\vec{x}_\Sigma$ and $\vec{y}_\Sigma$ and satisfy appropriate orthogonalities, and similarly make $\vec{x}_2$ and $\vec{y}_2$ be coplanar with $\vec{x}_\Pi$ and $\vec{y}_\Pi$ and satisfy appropriate orthogonalities. Explicitly, this choice is
\begin{equation}\label{eqn: x1}
\vec{x}_1 = \vec{x}_\Sigma - \frac{\vec{x}_\Sigma \cdot \vec{x}_\Sigma}{\vec{x}_\Sigma \cdot \vec{y}_\Sigma} \vec{y}_\Sigma
\end{equation}
and so on. It is easy to check that Eq.~\eqref{eqn: x1} is orthogonal to $\vec{x}_\Sigma$. We can then make $\vec{x}_{\mu>2}$ and $\vec{y}_{\mu>2}$ as orthogonal to these two planes.

\subsection{All the $\vec{x}_{\mu\neq\Sigma,\Pi}$ are mutually orthogonal, and analogously for $\vec{y}_\mu$.}
The Bogoliubov operators $\hat b_{\mu\neq\Sigma,\Pi}$ are independent of each other, i.e., $[\hat b_\mu, \hat b_\nu] = 0$ and $[\hat b_\mu, \hat b_\nu\+] = \delta_{\mu\nu}$. This leads to $\vec{x}_\mu\cdot\vec{y}_\nu = \frac{1}{2}\delta_{\mu\nu}$. A simple choice to satisfy these is to set $\vec{x}_{\mu > 2} = \vec{y}_{\mu > 2}$, and choose $\vec{x}_{\mu > 2}$ mutually orthogonal to each other. Since we already chose $\vec{x}_{\mu > 2}$ as orthogonal to the planes of $\vec{x}_1$ and $\vec{x}_2$, therefore all the $\vec{x}_\mu$ are orthogonal to each other. Similarly, all the $\vec{y}_\mu$ are orthogonal to each other.

\section{Schwinger bosons for the multilevel system}\label{app: Schwinger bosons}
For the basis chosen in Eq.~\eqref{eqn: schwinger bosons defn}, the $\vec{x}_\alpha$ and $\vec{y}_\alpha$ vectors are
\begin{align}\label{eqn: x and y}
 \vec{x}_\Sigma = &\left(0,0,0,0, \sqrt{\frac{2}{15}}\sin\frac{\beta}{2}-\frac{\cos(\beta/2)}{\sqrt{10}}, 0, -\frac{\sin(\beta/2)}{\sqrt{10}} \right)
, \nonumber\\
 \vec{y}_\Sigma = &\left(0, \frac{(-3\cos\beta + \sqrt{3}\sin\beta)\left(3\sin\sqrt{\frac{3}{5}}\theta_0 - \sin\frac{\theta_0}{\sqrt{15}} \right)}{12\sqrt{10}}, 0,0, \frac{9\cos\left(\sqrt{\frac{3}{5}}\theta_0\right)\left(\sqrt{3}\sin\frac{\beta}{2}-\cos\frac{\beta}{2}\right)-\cos\frac{\theta_0}{\sqrt{15}}\left(\sqrt{3}\sin\frac{\beta}{2}+3\cos\frac{\beta}{2}\right)}{12\sqrt{10}},
 \right.\nonumber\\ &\left.
 0, \frac{\cos\left(\sqrt{\frac{3}{5}}\theta_0\right)\left(\sqrt{3}\cos\frac{\beta}{2}-3\sin\frac{\beta}{2}\right)-\cos\frac{\theta_0}{\sqrt{15}}\left(\sqrt{3}\cos\frac{\beta}{2}+\sin\frac{\beta}{2}\right)}{4\sqrt{10}} \right)
 , \nonumber\\
 \vec{x}_\Pi = &\left( 
 -\frac{\cos\frac{\beta}{2}\sin\frac{2\theta_0}{\sqrt{15}}}{\sqrt{10}}, 0, \frac{\sin\frac{\beta}{2}\sin\frac{2\theta_0}{\sqrt{15}}}{\sqrt{10}}, -\frac{\sqrt{3}\cos\frac{\beta}{2}\cos^2\frac{\theta_0}{\sqrt{15}}+\sin\frac{\beta}{2}\sin^2\frac{\theta_0}{\sqrt{15}} }{\sqrt{10}}, 0, \frac{\frac{1}{\sqrt{12}}\left(3\cos\frac{2\theta_0}{\sqrt{15}}-1\right)-\cos\frac{\beta}{2}\sin^2\frac{\theta_0}{\sqrt{15}}}{6\sqrt{10}}, 0
 \right), \nonumber\\
 \vec{y}_\Pi = &\left( 
 -\frac{\sin\frac{\theta_0}{\sqrt{15}}\left(2\sin\frac{\beta}{2}-\sqrt{3}\cos\frac{\beta}{2}\right)}{\sqrt{30}}, 0, -\frac{1}{\sqrt{10}}\sin\frac{\theta_0}{\sqrt{15}}\sin\frac{\beta}{2}, -\sqrt{\frac{3}{10}}\cos\frac{\beta}{2}\cos\frac{\theta_0}{\sqrt{15}}, 0, \frac{1}{\sqrt{30}}\sin\frac{\beta}{2}\cos\frac{\theta_0}{\sqrt{15}}, 0
 \right).
\end{align}

Modifying the definition of the Schwinger c-bosons in Eq.~\eqref{eqn: schwinger bosons defn} would lead to a SU($\ell-1$) rotation on (the complex vectors) $\vec{x}_\alpha$ and $\vec{y}_\alpha$ in Eq.~\eqref{eqn: x and y}, which would leave the dot products such as $\vec{x}_\alpha \cdot \vec{y}_\alpha$ invariant. All physically relevant quantities such as criticality and squeezing depend only on these SU($\ell-1$) invariant constants.

\section{Finite-size corrections to the squeezing} \label{app: finite size correction}
The squeezed quadratures near the critical points are approximately $\hat X^c_{\alpha,\rm sq} = \frac{\sum_\mu x_{\alpha,\mu} \hat X^c_\mu}{\| \vec{x}_\alpha\|} \propto \hat X^b_\alpha$ and $\hat Y^c_{\alpha,\rm sq} = \frac{\sum_\mu y_{\alpha,\mu} \hat Y^c_\mu}{\| \vec{y}_\alpha\|} \propto \hat Y^b_\alpha$. The squeezing in these variables will be limited by finite-size effects, which can be captured by including the $O(1)$ terms in the master equation. Here, we will calculate the finite-size corrections, and the best squeezing reachable, by calculating and solving the steady-state value of $\braket{(\hat X^c_{\Sigma,\rm sq})^2}$. The calculations of the finite-size corrections to $\braket{(\hat X^c_{\Pi,\rm sq})^2}$ and $\braket{(\hat Y^c_{\alpha,\rm sq})^2}$ are similar.

We separate the jump operators into their leading and sub-leading terms, $\hat{\mathscr{D}}^-_\alpha = \sqrt{N}\hat{L}_{1,\alpha} + \underbrace{\hat{L}_{2,\alpha}}$, and expand the master equation
$\partial_t \braket{(\hat X^c_{\Sigma,\rm sq})^2} = \frac{\hbar\Gamma}{2}\sum_\alpha \braket{\hat{\mathscr{D}}^+_\alpha [(\hat X^c_{\Sigma,\rm sq})^2,\hat{\mathscr{D}}^-_\alpha] + {\rm H.c.}}$.
It can be shown that $\braket{\hat L_{1,\alpha}\+ [(\hat X^c_{\Sigma,\rm sq})^2,\underbrace{\hat L_{2,\alpha}}]} = 0$ and $[(\hat X^c_{\Sigma,\rm sq})^2,\underbrace{\hat L_{2,\Pi}}] = 0$. 
Expanding the remaining terms in the master equation gives
\begin{align}
&\partial_t \braket{\hat X^c_{\Sigma,\rm sq})^2} = \frac{ N\Gamma (\vx_\Sigma \cdot \vy_\Sigma)^2 }{\| \vec{x}_\Sigma \|^2} - 2N\Gamma (\vx_\Sigma \cdot \vy_\Sigma) \braket{(\hat X^c_{\Sigma,\rm sq})^2} + \underbrace{\frac{\Gamma}{8\| \vec{x}_\Sigma \|^2} \sum_\alpha \left\langle (\hvX^c -i\hvY^c)\cdot \lrvec{g}\+_\alpha \cdot (\hvX^c + i\hvY^c) \times \right.} \nonumber\\ 
& \underbrace{
\left( (\vx_\Sigma \cdot \hvX^c) (\vx_\Sigma \cdot \lrvec{g}_\alpha \cdot (\hvX^c + i\hvY^c)) - (\vx_\Sigma \cdot \hvX^c) (\vx_\Sigma \cdot \lrvec{g}^T_\alpha \cdot (\hvX^c - i\hvY^c)) + (\vx_\Sigma \cdot \lrvec{g}_\alpha \cdot (\hvX^c + i\hvY^c))(\vx_\Sigma \cdot \hvX^c) \right.}
 \nonumber\\ & \underbrace{\left. \left.
- (\vx_\Sigma \cdot \lrvec{g}^T_\alpha \cdot (\hvX^c - i\hvY^c))(\vx_\Sigma \cdot \hvX^c) \right) + {\rm H.c.} \right\rangle}_{\rm finite\ size}
\end{align}
where we have used shorthand notation $\vec{v}\cdot \hvX^c = \sum_{\mu>0} v_\mu \hat X^c_\mu$ and $\vec{v}\cdot\lrvec{g}_\alpha\cdot\hvX = \sum_{\mu\nu>0} v_\mu g_{\alpha,\mu\nu} \hat X^c_\nu$.

The steady-state value of $\braket{\hat X^c_{\Sigma,\rm sq})^2}$, obtained by setting its derivative to zero, is
\begin{align}\label{eqn: finite size correction 2}
&\braket{\hat X^c_{\Sigma,\rm sq})^2}_{\rm ss} = \frac{ \vx_\Sigma \cdot \vy_\Sigma }{2\|\vec{x}_\Sigma\|^2} + \underbrace{\frac{1}{16N\| \vec{x}_\Sigma \|^2(\vx_\Sigma\cdot\vy_\Sigma)} \sum_\alpha \left\langle (\hvX^c -i\hvY^c)\cdot \lrvec{g}\+_\alpha \cdot (\hvX^c + i\hvY^c) \times \right.} \nonumber\\ 
& \underbrace{
\left( (\vx_\Sigma \cdot \hvX^c) (\vx_\Sigma \cdot \lrvec{g}_\alpha \cdot (\hvX^c + i\hvY^c)) - (\vx_\Sigma \cdot \hvX^c) (\vx_\Sigma \cdot \lrvec{g}^T_\alpha \cdot (\hvX^c - i\hvY^c)) + (\vx_\Sigma \cdot \lrvec{g}_\alpha \cdot (\hvX^c + i\hvY^c))(\vx_\Sigma \cdot \hvX^c) \right.}
 \nonumber\\ & \underbrace{\left. \left.
- (\vx_\Sigma \cdot \lrvec{g}^T_\alpha \cdot (\hvX^c - i\hvY^c))(\vx_\Sigma \cdot \hvX^c) \right) + {\rm H.c.} \right\rangle_{\rm ss}}_{\rm finite\ size}
\end{align}
The expectation value in Eq.~\eqref{eqn: finite size correction 2} is calculated in the steady state $\rho_{\rm ss}$. It can be somewhat simplified using the fact that $\rho_{\rm ss}$ is the vacuum of $\hb_\Sigma$ and $\hb_\Pi$, which means that $(\vx_\Sigma \cdot \hvX^c + i\vy_\Sigma \cdot \hvY^c)\rho_{\rm ss} = (\vx_\Pi \cdot \hvX^c + i\vy_\Pi \cdot \hvY^c)\rho_{\rm ss} = 0$ and $\braket{X_j Y_k}_{\rm ss} = i\delta_{jk}/2$. This simplification yields
\begin{align}\label{eqn: finite size correction 3}
&\braket{\hat X^c_{\Sigma,\rm sq})^2}_{\rm ss} = \frac{ \vx_\Sigma \cdot \vy_\Sigma }{2\|\vec{x}_\Sigma\|^2} - \underbrace{ \frac{1}{16N(\vx_\Sigma \cdot \vy_\Sigma)\|\vec{x}_\Sigma\|^2} \times \sum_\alpha \left( \right.} \nonumber\\
&\underbrace{ \left\langle (\vx_\Sigma\cdot\lrvec{g}_\alpha^S\cdot\hvY^c)^2 + (\vx_\Sigma\cdot\lrvec{g}_\alpha^A\cdot\hvX^c)^2 + (\vx_\Sigma\cdot\lrvec{g}_\alpha^A\cdot\hvX^c)(\vy_\Sigma\cdot\lrvec{g}_\alpha^S\cdot\hvX^c) + (\vy_\Sigma\cdot\lrvec{g}_\alpha^A\cdot\hvY^c)(\vx_\Sigma\cdot\lrvec{g}_\alpha^S\cdot\hvY^c) \right.}
\nonumber\\ & \underbrace{\left.\left. + \vx_\Sigma\cdot\lrvec{g}_\alpha^A\cdot\vy_\Sigma( \hvX^c\cdot\lrvec{g}_\alpha\cdot\hvX^c + \hvY^c\cdot\lrvec{g}_\alpha
\cdot\hvY^c) \right\rangle_{\rm ss} 
+
\frac{2 \vx_\Sigma\cdot\lrvec{g}_\alpha^S\cdot\lrvec{g}_\alpha^A\cdot\vx_\Sigma + \vy_\Sigma\cdot\lrvec{g}_\alpha^A\cdot\lrvec{g}_\alpha^A\cdot\vx_\Sigma - \vy_\Sigma\cdot\lrvec{g}_\alpha^S\cdot\lrvec{g}_\alpha^S\cdot\vx_\Sigma }{2}
\right)}_{\rm finite\ size}
\end{align}
where $\lrvec{g}_\alpha^S = \lrvec{g}_\alpha + \lrvec{g}_\alpha^T$ and $\lrvec{g}_\alpha^A = \lrvec{g}_\alpha - \lrvec{g}_\alpha^T$.

The dominant contribution to the $O(1/N)$ terms comes from the antisqueezed quadrature. Keeping only this dominant contribution, the value of the steady-state squeezing is approximately
\begin{align}
&\braket{\hat X^c_{\Sigma,\rm sq})^2}_{\rm ss} \simeq \frac{ \vx_\Sigma \cdot \vy_\Sigma }{2\|\vec{x}_\Sigma\|^2} - \underbrace{
\frac{\|\vec{y}_\Sigma\|^2}{16N(\vx_\Sigma \cdot \vy_\Sigma)^3} \sum_\alpha \left( \frac{(\vx_\Sigma\cdot\lrvec{g}_\alpha^S\cdot\vx_\Sigma)^2}{\|\vx_\Sigma\|^2} 
\right. }
\nonumber\\
&\underbrace{ + \left. 
\frac{(\vx_\Sigma\cdot\lrvec{g}_\alpha^A\cdot\vy_\Sigma)^2}{\|\vy_\Sigma\|^2} + \frac{\vx_\Sigma\cdot\lrvec{g}_\alpha^A\cdot\vy_\Sigma}{2} \left( 3\frac{\vy_\Sigma\cdot\lrvec{g}_\alpha^S\cdot\vy_\Sigma}{\|\vy_\Sigma\|^2} - \frac{\vx_\Sigma\cdot\lrvec{g}_\alpha^S\cdot\vx_\Sigma}{\|\vx_\Sigma\|^2}\right) \right)}_{\rm finite\ size}.
\end{align}
The coefficient of the second term, $\underbrace{\frac{1}{N(\vx_\Sigma \cdot \vy_\Sigma)^3}}$, tends to a constant value as the critical point is approached. This constant value is denoted $A$ in Eq.~\eqref{eqn: multilevel squeezing near critical pt}.

\end{widetext}
\end{document}